\documentclass[journal]{IEEEtran}

\ifCLASSOPTIONcompsoc
\usepackage[nocompress]{cite}
\else
\usepackage{cite}
\fi

\usepackage[OT1]{fontenc} 
\usepackage[numbers,sort&compress]{natbib}
\usepackage[cmex10]{amsmath}
\usepackage{amssymb}
\usepackage{bm}
\usepackage{graphicx}
\usepackage{color}

\usepackage{subfigure}
\usepackage{tabularx}
\usepackage{arydshln}
\usepackage{mathtools}
\usepackage{multirow}
\usepackage{algorithm,algorithmic}
\usepackage{url}
\usepackage{braket}
\usepackage{comment}
\usepackage{hyperref}
\usepackage[absolute,overlay]{textpos}

\newcommand{\inbb}[2]{\in \mathbb{#1}^{#2}}
\def\b{\mathbf{b}}

\def\etal{\textit{et al.}~}

\def\G{\mathbf{G}}

\def\I{\mathbf{I}}

\def\U{\mathbf{U}}

\def\0{\mathbf{0}}
\def\1{\mathbf{1}}

\def\({\left(}
\def\){\right)}
\def\[{\left[}
\def\]{\right]}

\makeatletter
\def\ps@IEEEtitlepagestyle{%
  \def\@oddfoot{\mycopyrightnotice}%
  \def\@oddhead{\hbox{}\@IEEEheaderstyle\leftmark\hfil\thepage}\relax
  \def\@evenhead{\@IEEEheaderstyle\thepage\hfil\leftmark\hbox{}}\relax
  \def\@evenfoot{}%
}
\def\mycopyrightnotice{%
  \begin{minipage}{\textwidth}
  \centering \scriptsize
\textcopyright 2023 IEEE. Personal use of this material is permitted.
  Permission from IEEE must be obtained for all other uses, in any current or future
  media, including reprinting/republishing this material for advertising or promotional
  purposes, creating new collective works, for resale or redistribution to servers or
  lists, or reuse of any copyrighted component of this work in other works.
  DOI: \href{https://doi.org/10.1109/TCOMM.2023.3244924}{10.1109/TCOMM.2023.3244924}
  \end{minipage}
}
\makeatother
\begin{document}
\title{Quantum Algorithm for Higher-Order Unconstrained Binary Optimization and MIMO Maximum Likelihood Detection}
\author{Masaya~Norimoto,~\IEEEmembership{Graduate Student Member,~IEEE},
Ryuhei~Mori,~\IEEEmembership{Non~Member,~IEEE}, and Naoki~Ishikawa,~\IEEEmembership{Senior~Member,~IEEE}.\thanks{M.~Norimoto and N.~Ishikawa are with the Faculty of Engineering, Yokohama National University, Kanagawa 240-8501, Japan (e-mail: ishikawa-naoki-fr@ynu.ac.jp).
R. Mori is with the Department of Mathematical and Computing Sciences, School of Computing, Tokyo Institute of Technology, Tokyo 152-8500, Japan (e-mail: mori@c.titech.ac.jp).
This research was partially supported by the Japan Society for the Promotion of Science (JSPS) KAKENHI (Grant Number 22H01484).}}

%FOR ARXIV
%\markboth{Accepted for publication in IEEE Transactions on Communications. This is the author's version which has not been fully edited and content may change prior to final publication. Citation information: DOI 10.1109/TCOMM.2023.3244924}

\maketitle
\TPshowboxesfalse
\begin{textblock*}{\textwidth}(45pt,10pt)
\footnotesize
\centering
Accepted for publication in IEEE Transactions on Communications. This is the author's version which has not been fully edited and content may change prior to final publication. Citation information: DOI 10.1109/TCOMM.2023.3244924
\end{textblock*}
%FOR ARXIV

\begin{abstract}
In this paper, we propose a quantum algorithm that supports a real-valued higher-order unconstrained binary optimization (HUBO) problem. This algorithm is based on the Grover adaptive search that originally supported HUBO with integer coefficients. Next, as an application example, we formulate multiple-input multiple-output maximum likelihood detection as a HUBO problem with real-valued coefficients, where we use the Gray-coded bit-to-symbol mapping specified in the 5G standard. The proposed approach allows us to construct an efficient quantum circuit for the detection problem and to analyze specific numbers of required qubits and quantum gates, whereas other conventional studies have assumed that such a circuit is feasible as a quantum oracle. To further accelerate the quantum algorithm, we also derive a probability distribution of the objective function value and determine a unique threshold to sample better states. Assuming a future fault-tolerant quantum computing, our proposed algorithm has the potential for significantly reducing query complexity in the classical domain and providing a quadratic speedup in the quantum domain.
\end{abstract}


\IEEEpeerreviewmaketitle

\section{Introduction\label{sec:intro}}
\IEEEPARstart{M}{arconi} invented a practical long-range wireless system in 1895.
Since then, driven by its intense demand, wireless communication has continued to become more sophisticated as if there were no limits. The limit of communication throughput is known as the Shannon capacity, which is constrained by the bandwidth, the signal-to-noise ratio (SNR), and the numbers of transmit and receive antennas for multiple-input multiple-output (MIMO) scenarios.
Clearly, there are physical limits on bandwidth, SNR, and the number of antennas.
The forward error correction techniques such as the low-density parity-check code (LDPC) and polar code can achieve near-capacity performance efficiently, but under a certain energy constraint, their performance is constrained by semiconductor miniaturization limits.
Marconi mentioned that \textit{it is dangerous to put limits on wireless}. However, wireless communication will reach its physical limits in the near future.

After the eventual end of Moore's law, from a long-term perspective, we must rely on a different computing paradigm, and quantum computing in particular is considered to be promising.
Since it is impossible to simulate a quantum computer in an efficient manner on a classical computer, quantum computers offer an essential speed advantage over classical computers \cite{nielsen2010quantum}.
Specifically, Shor's algorithm \cite{shor1994algorithms} factors an $n$-bit integer with the complexity $O(n^2 \log n \log \log n)$, while the best classical algorithm requires $\exp(\Theta(n^{1/3} \log^{2/3} n))$ operations \cite{nielsen2010quantum},\footnote{$O(\cdot)$ denotes the big-$O$ notation, while $\Theta(\cdot)$ denotes the big-$\Theta$ notation \cite{knuth1976big}.} which is an \textit{exponential speedup}.
Grover's algorithm \cite{grover1996fast} finds a specific element from a database of unsorted $N$ elements with the query complexity $O(\sqrt{N})$, while the classic exhaustive search requires $O(N)$ evaluations, which is a \textit{quadratic speedup}.
Note that both long-term algorithms assume the realization of fault-tolerant quantum computing (FTQC), which is not yet realized with current technology.

Grover's algorithm has been extended to support binary optimization problems.
The pioneering algorithm, \textit{Grover adaptive search} (GAS) \cite{bulger2003implementing}, requires a complex quantum circuit to evaluate an objective function.
For example, an $m$-qubit register requires $2m-1$ Toffoli gates to perform quantum addition \cite{gidney2018halving}, which is still expensive.
To solve this issue, Gilliam \etal used a concept of quantum dictionary and allowed for the efficient representation of an arbitrary polynomial function, including quadratic and higher-order terms \cite{gilliam2020optimizing,gilliam2021grover}.
This efficient representation improved the feasibility of GAS, and in \cite{gilliam2021grover}, a quadratic unconstrained binary optimization (QUBO) problem with integer coefficients was solved on a real-world quantum computer with 32 qubits.
Unlike the quantum annealing (QA) \cite{kadowaki1998quantum}, the GAS proposed by Gilliam \etal is innovative in that it supports a higher-order unconstrained binary optimization (HUBO) problem with integer coefficients, which cannot be solved efficiently with state-of-the-art mathematical programming solvers on a classical computer such as CPLEX\footnote{\url{https://www.ibm.com/analytics/cplex-optimizer}}.

In designing wireless systems, the trade-off between performance and complexity is, in general, a source of concern for engineers and researchers.
For example, low-complexity MIMO detectors and polar decoders inevitably involve the penalty of lower performance, and complexity is sacrificed to achieve optimal performance.
In this situation, the potential for quantum speedup has inspired those who dream of striking the fundamental trade-off and achieving the optimal performance with reduced complexity.
A pioneering attempt in wireless communications was provided in \cite{botsinis2013quantum} by Botsinis \etal who demonstrated the potential of quantum algorithms to reduce the complexity involved in maximum likelihood detection (MLD).
Specifically, they used the Grover-type algorithms, such as Boyer--Brassard--H\o yer--Tapp (BBHT) \cite{boyer1998tight} and D\" urr--H\o yer (DH) search algorithms \cite{durr1999quantum}, for performing MLD of data symbols on a quantum computer \cite{botsinis2014fixedcomplexity}.
Subsequently, a number of important studies have shown promising results \cite{botsinis2014fixedcomplexity,botsinis2014lowcomplexity,botsinis2015iterative,botsinis2015noncoherent,ye2019quantum,alanis2018quantumsearchaided,alanis2018quantumaided,botsinis2017quantumassisted}. However, in those studies, it was assumed that an ideal quantum circuit to evaluate the objective function is feasible as a quantum oracle, which will be detailed in Section~\ref{sec:rel}.
For more information on quantum-assisted wireless communications, a comprehensive survey can be found in \cite{botsinis2017coherent,botsinis2019quantum}.

Against this background, we propose a quantum algorithm that supports a HUBO problem with real-valued coefficients.
Then, as a first step toward breaking the trade-off between performance and complexity, we formulate the MIMO MLD as a real-valued HUBO problem and verify the potential of quadratic speedup.
The major contributions of this paper are organized as follows.
\begin{enumerate}
    \item While the conventional GAS \cite{gilliam2021grover} supports HUBO with integer coefficients, we modify the quantum algorithm to handle real-valued coefficients. This allows us to solve a HUBO problem even if the objective function contains real-valued coefficients, which is achieved at the cost of one more query in the classical domain (CD).\footnote{The definition of query complexity in CD is detailed in Section~\ref{subsec:metrics}, which is the same as the conventional study \cite{botsinis2014fixedcomplexity}.}
    \item As an application example, we formulate the objective function of MIMO MLD as a real-valued HUBO problem. This formulation is not a straightforward task because the objective function contains complex-valued random variables and a Frobenius norm calculation. This new formulation allows us to analyze specific numbers of qubits and quantum gates required in the constructed quantum circuits, which has been overlooked in conventional studies.
    \item We clarify the probability distribution of the objective function value and determine the threshold used inside GAS more efficiently. Then, we demonstrate that the proposed threshold further accelerates the convergence of GAS to the optimal solution. 
\end{enumerate}

It is important to note that quantum circuits are sensitive to noise \cite{fujii2016noise}, and industrial applications require decades of effort and challenge. The noise induces quantum errors, and quantum error-correcting codes must be used to perform reliable arithmetic on a quantum computer. For example, if we use the surface code with code distance 27, which is one of the quantum error-correcting codes, a logical qubit requires 1568 physical qubits to correct errors \cite{gidney2021how}. This indicates that even a simple quantum circuit with fewer qubits, e.g., as in Fig.~\ref{fig:circuit-int}, may require many more physical qubits.
%d = 27, 2*(d + 1)^2
Since this limitation is beyond the scope of our contributions, we assume the realization of future FTQC as in the conventional studies \cite{shor1994algorithms,grover1996fast,bulger2003implementing,gilliam2020optimizing,gilliam2021grover,botsinis2013quantum,boyer1998tight,durr1999quantum,botsinis2014fixedcomplexity,botsinis2014lowcomplexity,botsinis2015iterative,botsinis2015noncoherent,ye2019quantum,alanis2018quantumsearchaided,alanis2018quantumaided,botsinis2017quantumassisted,botsinis2017coherent,botsinis2019quantum,ishikawa2021quantum}. In fact, IBM's roadmap for quantum computers is to achieve 4000 qubits by 2025 \cite{jay2022ibm}. In the subsequent years, they expect 10000 to 100000 qubits, enabling quantum error corrections.
Additionally, it was proved in \cite{stilckfranca2021limitations} that quantum advantages are unlikely for optimization on a noisy intermediate-scale quantum device.
Therefore, we focus on long-term algorithms assuming FTQC in this paper.

The remainder of this paper is organized as follows.
Section~\ref{sec:rel} is a review of important related works, while in Section~\ref{sec:gas}, we introduce the conventional GAS and its modification to support real-valued coefficients.
In Section~\ref{sec:mld}, a method to solve MIMO MLD on a quantum computer is proposed, and algebraic and numerical evaluations are given in Section~\ref{sec:comp}. Finally, in Section~\ref{sec:conc}, we conclude this paper.
\begin{table}[tb]
	\centering
	\caption{List of important mathematical symbols\label{table:sym}}
	\begin{tabular}{lll}
	    $\mathbb{B}$ & & Binary numbers \\
	    $\mathbb{R}$ & & Real numbers \\
	    $\mathbb{C}$ & & Complex numbers \\
	    $\mathbb{Z}$ & & Integers \\
	    $N_{\mathrm{t}}$ & $\inbb{Z}{}$ & Number of transmit antennas\\
		$N_{\mathrm{r}}$ & $\inbb{Z}{}$ & Number of receive antennas\\
		$L_{\mathrm{c}}$ & $\inbb{Z}{}$ & Modulation order (constellation size)\\
		$\sigma^2$ & $\inbb {R}{}$ & Noise variance \\
		$\gamma$ & $\inbb {R}{}$ & Signal-to-noise ratio \\
		$E(\cdot)$ & $\inbb{Z}{}$ & Objective function \\
		$n$ & $\inbb{Z}{}$ & Number of binary variables $=$ transmission rate \\
		$m$ & $\inbb{Z}{}$ & Number of qubits required to encode $E(\cdot)$ \\
		$i$ & $\inbb{Z}{}$ & Index of GAS iterations \\
		$y, y_i$ & $\inbb{Z}{}$ & Threshold that is adaptively updated by GAS\\
		$L, L_i$ & $\inbb{Z}{}$ & Number of Grover operators\\
		$P$ & $\inbb{R}{}$ & Probability that controls the proposed threshold\\
		$\mathbf{b}, \mathbf{b}_i$ & $\inbb {B}{n}$ & Binary variables, or data bits \\
		$\mathbf{s}$ & $\inbb {C}{N_{\mathrm{t}}\times 1}$ & Data symbols, each symbol is denoted by $s_t$ \\
		$\mathbf{r}$ & $\inbb {C}{N_{\mathrm{r}}\times 1}$ & Received symbols, each symbol is denoted by $r_u$ \\
		$\mathbf{H}_{\mathrm{c}}$ & $\inbb {C}{N_{\mathrm{r}}\times N_{\mathrm{t}}}$ & Channel coefficients, $h_{ut}$ \\
		$\mathbf{v}$ & $\inbb {C}{N_{\mathrm{r}}\times 1}$ & Additive white Gaussian noise, $v_u$ \\
	\end{tabular}
\end{table}
Italicized symbols represent scalar values, and bold symbols represent vectors and matrices.
Table~\ref{table:sym} summarizes a list of important mathematical symbols used in this paper.

\section{Related Works\label{sec:rel}}
Quantum computation has the potential to break through the fundamental trade-off between performance and complexity.
Hence, it has been applied to multi-user detection \cite{botsinis2013quantum,botsinis2014fixedcomplexity,botsinis2014lowcomplexity,botsinis2015iterative,ye2019quantum,kim2019leveraging,mondal2021ml}, multiple symbol differential detection \cite{botsinis2015noncoherent}, channel coding \cite{babar2020polar,matsumine2019channel}, wireless routing \cite{alanis2018quantumaided,alanis2018quantumsearchaided}, indoor localization \cite{botsinis2017quantumassisted}, intelligent reflecting surfaces \cite{ohyama2021intelligent}, and codeword optimization problem \cite{ishikawa2021quantum}.
In this section, we introduce important related works targeting detection problems in wireless communications.

\subsection{Multi-User Detection Using DH Algorithm \cite{botsinis2014fixedcomplexity}}
Botsinis \etal proposed a novel method of applying the DH algorithm to multi-user detection \cite{botsinis2014fixedcomplexity}, which is a detection problem for multi-user scenarios.
The original DH algorithm \cite{durr1999quantum} is terminated if the sum of the number of Grover iterations becomes greater than or equal to $22.5 \sqrt{N}$, where $N$ denotes the search space size.
By contrast, Botsinis \etal modified the algorithm to terminate early for an arbitrary number of queries smaller than $22.5 \sqrt{N}$.
Additionally, the modified algorithm calculates the output of a low-complexity detector, such as the zero-forcing (ZF) or minimum mean square error (MMSE) detector, and exploits the output as an initial value to sample better states.
Both contributions are innovative in that they accelerate the quantum algorithm more for a specific problem in wireless communications.

The objective function presented in \cite{botsinis2014fixedcomplexity} involves a Frobenius norm of complex-valued variables.
However, the quantum circuit that evaluates the norm is idealized as an oracle, and no specific construction method is considered.
Unlike in \cite{botsinis2014fixedcomplexity}, we consider specific quantum circuits and analyze their hardware and query complexities, which is the missing piece in the literature.

\subsection{MIMO MLD Using QA \cite{kim2019leveraging}}
Kim \etal formulated MIMO MLD as a QUBO problem and solved it using QA, the D-Wave 2000Q quantum annealer \cite{kim2019leveraging}.
Specifically, binary phase-shift keying (BPSK) and quadrature phase-shift keying (QPSK) symbols are represented as first-order functions with respect to information bits, while gray-coded 16 quadrature amplitude modulation (QAM) symbols are represented as second-order functions.
Since the objective function of MLD contains the squared norm, it may result in a higher-order function such as fourth, eighth, or higher, which is not supported by QA.
To solve this problem, Kim \etal used first-order functions that represent higher-order modulation, such as 16-QAM or 64-QAM, without the Gray coding.
Then, the objective function contains first- and second-order terms only.
To achieve performance equivalent to that of the Gray-coded case, the projection between before and after Gray coding is used on a classical computer.
That is, encoding at the transmitter and decoding at the receiver require additional steps.

Unlike in the above study \cite{kim2019leveraging} targeting QA, we directly handle the Gray-coded data symbols specified in the 5G standard owing to the proposed real-valued GAS that supports higher-order terms.
Our approach is capable of supporting any signal modulation, such as star-QAM and constellation shaping schemes, as long as data symbols can be represented as a function of information bits.

\subsection{MIMO MLD Using DH Algorithm \cite{mondal2021ml}}
Mondal \etal proposed a method to solve MIMO MLD using the DH algorithm \cite{mondal2021ml}.
Specifically, to improve the success probability of the algorithm, the uniform selection of the number of Grover operators, $L$, was modified to a random value from the Gamma distribution, leading to a better selection of $L$.
Here, the Gamma distribution depends on a scale parameter, and the scale parameter depends on the exact number of solutions to be marked.
Since the exact number of solutions varies dynamically depending on the threshold, the quantum counting algorithm \cite{brassard1998quantum} is crucial, as stated in \cite{mondal2021ml}.
Additionally, the concept of reducing the search space was verified.

As in \cite{botsinis2014fixedcomplexity}, a specific construction method for a quantum circuit is not considered in \cite{mondal2021ml}.
Herein, we determine a threshold in accordance with the distribution of objective function values, which is known in advance.

\section{Grover Adaptive Search (GAS) \label{sec:gas}}
GAS \cite{gilliam2021grover} supports binary optimization problems with integer coefficients, including QUBO and HUBO problems.
It requires $n$ qubits for $n$ binary variables $\b \in \mathbb{B}^{n}$ and $m$ qubits for encoding the objective function value $E(\b) \in \mathbb{Z}$, resulting in a circuit equipped with $n+m$ qubits.
Here, $E(\b)$ is an arbitrary polynomial function, such as $E(\b)=1+b_0-2b_1b_2$.
The classic exhaustive search requires $O(2^n)$ queries, while GAS requires $O(\sqrt{2^n})$ queries, which potentially provides a quadratic speedup.
GAS obtains a global minimum solution by amplifying the states in which the objective function value $E(\b)$ is smaller than the current threshold $y_i \in \mathbb{Z}$.
Here, $y_i$ is a temporal minimum and $i$ is an iteration count in CD.
We measure the quantum states and update the threshold, which is repeated until a termination condition is satisfied.

Before executing GAS, it is not a straightforward task to determine an appropriate number of qubits $m$.
The objective function value is expressed by the two's complement representation. 
This is because positive or negative can be identified simply by focusing on the beginning of $m$ qubits, and this representation simplifies the quantum circuit to the identification of the states of interest that should be amplified.
Let the objective function value or its coefficient be an integer $k$. Then, $m$ must satisfy \cite{gilliam2021grover}
\begin{align}
-2^{m-1}\leq k<2^{m-1}.\label{eq:mconstraint}
\end{align}
As the threshold $y_i$ is updated in each iteration of GAS, the calculated value may become $E(\b)-y_i$, which results in a smaller minimum value or a larger maximum value.
Thus, it is necessary to set a sufficient $m$ that might handle $E_{\mathrm{max}}$, $E_{\mathrm{min}}$, and $E_{\mathrm{max}} - E_{\mathrm{min}}$ without overflow, where $E_{\mathrm{max}}$ and $E_{\mathrm{min}}$ are the maximum and minimum of $E(\b)$, respectively.
For example, when we have $E_{\mathrm{max}}=8$ and $E_{\mathrm{min}}=-6$, the maximum of $E(\b)-y_i$ may become $E_{\mathrm{max}} - E_{\mathrm{min}} = 8 - (-6) = 14$, and $m = 5$ is sufficient to represent the values of $E(\b)-y_i$.

\subsection{Conventional GAS for Integer QUBO \cite{gilliam2021grover}}
We review a specific construction method for the quantum circuit used in GAS.
First, a state preparation operator $\mathbf{A}_{y_i}$ is constructed, in which an $n$-qubit input register is transformed into the equal superposition of all states and an $m$-qubit input register is used to represent the corresponding value $E(\b)-y_i$.
Taking the binary variable $\b$ as a binary number and converting it to a decimal number $b$, the state should be \cite{gilliam2021grover}
\begin{align}
\mathbf{A}_{y_i}\Ket{0}_n \Ket{0}_m=\frac{1}{\sqrt{2^n}}\sum_{b=0}^{2^n-1}\Ket{b}_n\Ket{E(b)-y_i}_m.
\end{align}
This operator $\mathbf{A}_{y_i}$ can be composed of the Hadamard gates $\mathbf{H}$, controlled unitary operators $\mathbf{U}_G(\theta)$, and the inverse quantum Fourier transform (IQFT). 
Let $k$ be a constant term in the objective function.
The noncontrolled unitary operator $\mathbf{U}_G(\theta)$ is defined such that \cite{gilliam2021grover}
\begin{align}
\label{Ug-theta}
\mathbf{U}_G(\theta)\mathbf{H}^{\otimes m}\Ket{0}_{m}=\frac{1}{\sqrt{2^m}}\sum^{2^m-1}_
{l=0}e^{jl\theta}\Ket{l}_{m},
\end{align}
where we have $\theta=2\pi k/2^m$.
That is, it is constructed by
\begin{align}
    \mathbf{U}_G(\theta) = \mathbf{R}(2^{m-1} \theta)
    \otimes \mathbf{R}(2^{m-2} \theta) \otimes \cdots \otimes \mathbf{R}(2^{0} \theta)
\end{align}
and the phase gate
\begin{align}
    \mathbf{R}(\theta) = \begin{bmatrix}
        1 & 0 \\ 0 & e^{j\theta}
    \end{bmatrix}.
\end{align}
Here, phase advance represents integer addition and phase delay represents subtraction.
Following \eqref{Ug-theta}, IQFT yields only one state that represents the original integer value of $k$.

\begin{figure}[tb]
	\centering
    \includegraphics[clip, scale=0.58]{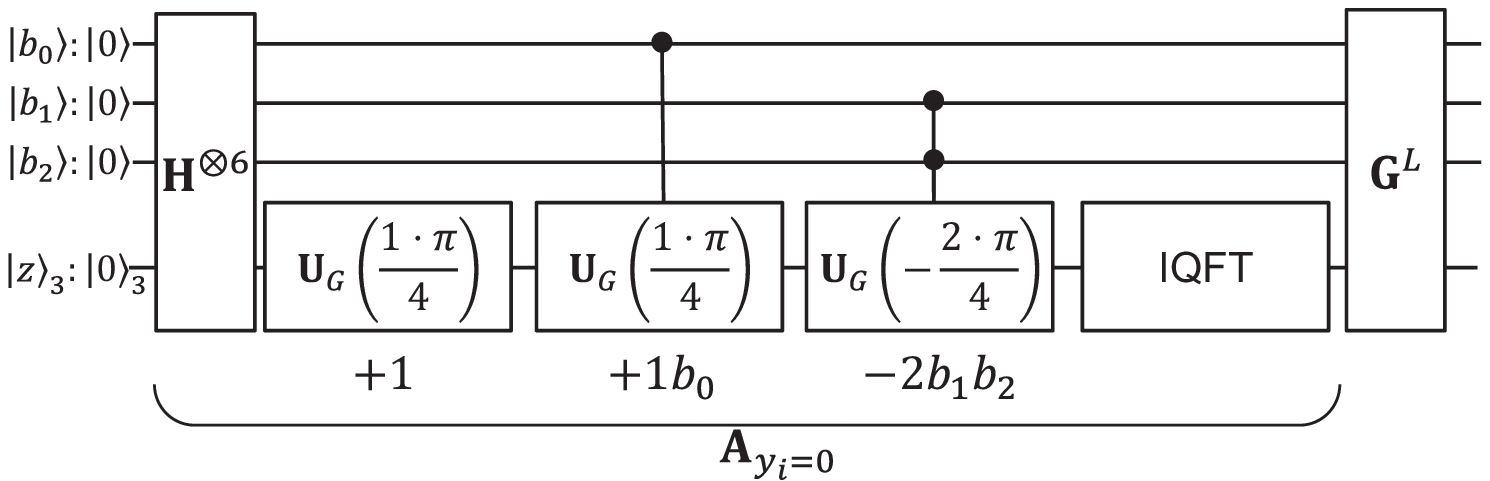}
	\caption{Quantum circuit corresponding to $E(\b)=1+b_0-2b_1b_2$. \label{fig:circuit-int}}
	
	\subfigure[$L=0$.]{
		\includegraphics[clip, scale=0.65]{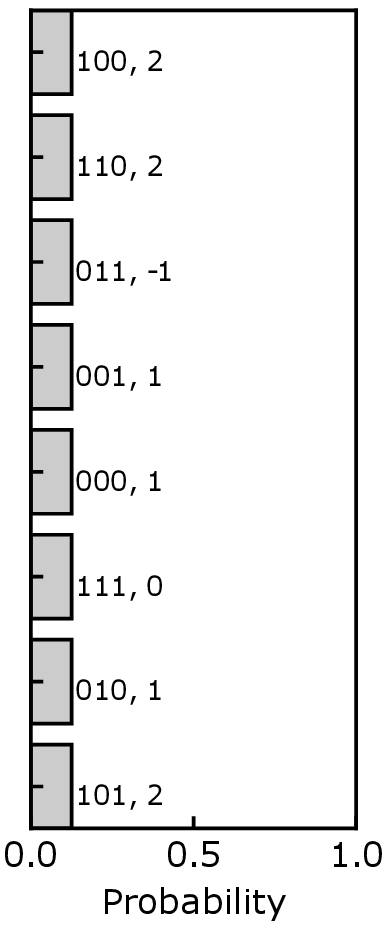}
	}
	\subfigure[$L=1$.]{
		\includegraphics[clip, scale=0.65]{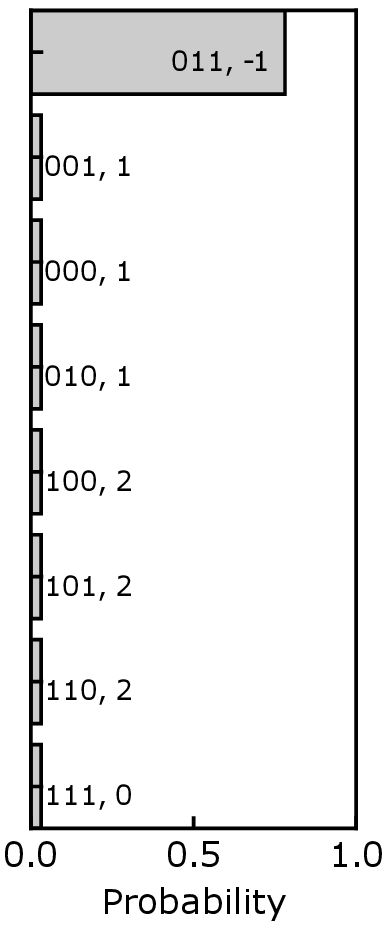}
	}
	\subfigure[$L=2$.]{
		\includegraphics[clip, scale=0.65]{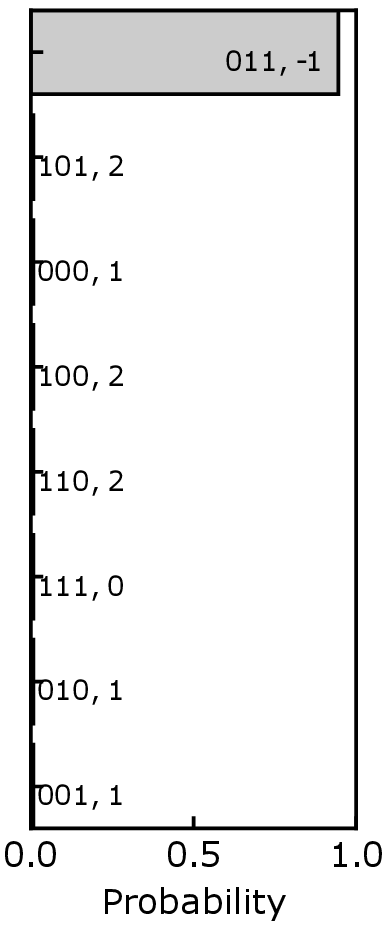}
	}
	\caption{Output probabilities of the circuit shown in Fig.~\ref{fig:circuit-int}\label{fig:circuit-int-out}}
\end{figure}

The interaction between a binary variable and a coefficient can be represented by a controlled qubit.
Similarly, the interaction between binary variables can be represented by controlled qubits on a register $\Ket{b}_{n}$.
As exemplified in Fig.~\ref{fig:circuit-int},
the constant term $+1$ corresponds to $\mathbf{U}_G\left( \pi / 4 \right)$,
the term $+1 b_0$ corresponds to controlled $\mathbf{U}_G\left( \pi / 4 \right)$,
and the term $-2 b_1 b_2$ corresponds to controlled $\mathbf{U}_G\left(- 2 \pi / 4 \right)$.
Likewise, higher-order terms, such as third or fourth order, can be represented by increasing the number of controlled qubits.

In the classic Grover search \cite{grover1996fast}, an oracle operator $\mathbf{O}$ identifies the states of interest and inverts the phases of these states.
Only the inverted states are amplified by the Grover operator. 
The operator $\mathbf{A}_{y_i}$ above calculates the values $E(\b)-y_i$ for all $2^n$ states in parallel.
Here, states that are better than the current threshold $y_i$, i.e., states that satisfy $E(\b)-y_i<0$, should be marked to find the minimum solution.
Since the calculated values are represented by the two's complement, we can identify the negative states by focusing only on the beginning of the $m$ qubits, and $\mathbf{O}$ can be constructed by applying the Z-gate only to that qubit. 

Let the Grover diffusion operator be $\mathbf{D}$ of \cite{grover1996fast}
\begin{align}
    \label{eq:D-def}
    D_{i,j} =
    \begin{cases}
        0 & (i \ne j)\\
        1 & (i =j= 0) \\
        -1 & (i =j\ne 0)
    \end{cases}.
\end{align}
The Grover operator is finally constructed by $\mathbf{G}=\mathbf{A}_{y_i}\mathbf{D}\mathbf{A}_{y_i}^{\mathrm{H}} \mathbf{O}$, and we evaluate $\mathbf{G}^L \mathbf{A}_{y_i} \Ket{0}_{n+m}$, which will maximize the amplitudes of the states of interest.
The ideal $L$ that successfully maximizes the amplitude is given by \cite{brassard2002quantum}
\begin{align}
\label{opt-g}
L_{\mathrm{opt}}=\left\lfloor \frac{\pi}{4}\sqrt{\frac{N}{N_{\mathrm{s}}}} \right\rfloor,
\end{align}
where $N$ denotes the search space size, $2^n$, and $N_{\mathrm{s}}$ denotes the number of solutions. Let $\mathbf{b}$ be the bit sequence and $y$ be the objective function value obtained by the evaluation.

From \eqref{opt-g}, the query complexity of GAS can be derived as $O(\sqrt{2^n})$ \cite{gilliam2021grover} in the quantum domain (QD), which is the total number of Grover operators.
Since $N_s$, the number of states better than the current threshold, is unknown in advance, $L$ is typically drawn from a uniform distribution ranging from $0$ to a specific value that increases by a factor of $\lambda = 8/7$ at each iteration.
GAS is terminated if the sum of the Grover operators is greater than $22.5\sqrt{2^n}$, which is the same as the conventional DH algorithm.
In the Qiskit implementation, the number of times no improvement is observed is also considered as one of the termination conditions.
Overall, GAS is summarized in Algorithm~\ref{alg:pro}.
\begin{algorithm}[tb]
    \caption{Conventional GAS designed for integer coefficients~\cite{gilliam2021grover}.\label{alg:pro}}
    \begin{algorithmic}[1]
        \renewcommand{\algorithmicrequire}{\textbf{Input:}}
        \renewcommand{\algorithmicensure}{\textbf{Output:}}
        \REQUIRE $E:\mathbb{B}^n\rightarrow\mathbb{Z}, \lambda=8/7$
        \ENSURE $\mathbf{b}$
        \STATE {Uniformly sample $\b_0 \in \mathbb{B}^n $ and set $y_0=E(\b_0)$}.
        
        \STATE {Set $k = 1$ and $i = 0$}.
        
        \REPEAT
        \STATE\hspace{\algorithmicindent}{Randomly select the rotation count $L_i$ from the set $\{0, 1, ..., \lceil k-1 \rceil$\}}.
        \STATE\hspace{\algorithmicindent}{Evaluate $\mathbf{G}^{L_i} \mathbf{A}_{y_i} \Ket{0}_{n+m}$, and obtain $\b$ and $y$}. \COMMENT{Grover search}
        \hspace{\algorithmicindent}\IF{$y<y_i$}
        \STATE\hspace{\algorithmicindent}{$\b_{i+1}=\b, y_{i+1}=y,$ and $k=1$}. \COMMENT{Improvement found}
        \hspace{\algorithmicindent}\ELSE{\STATE\hspace{\algorithmicindent}{$\b_{i+1}=\b_i, y_{i+1}=y_i,$ and $k=\min{\{\lambda k,\sqrt{2^n}}\}$}}. \COMMENT{No Improvement}
        \ENDIF
        \STATE{$i=i+1$}.
        \UNTIL{a termination condition is met}.
    \end{algorithmic} 
\end{algorithm}

As a specific example, Fig.~\ref{fig:circuit-int} shows a quantum circuit of GAS that tries to minimize the objective function $E(\b)=1+b_0-2b_1b_2$, where the threshold of $y_i = 0$ and $\mathbf{A}_{y_i=0}$ are considered for simplicity.
The upper $n=3$ qubits correspond to variables $\Ket{b_0}$, $\Ket{b_1}$, and $\Ket{b_2}$, and the lower $m=3$ qubits $\Ket{z}_3$ encode the calculated value.
The Hadamard gate at the beginning of $\mathbf{A}_{y_i=0}$ initializes the qubits and creates an equal superposition of all the possible states, $000000$ to $111111$.
The black circle in Fig.~\ref{fig:circuit-int} indicates a control qubit.
The unitary operator $\mathbf{U}_G(\theta)$ is applied if all the associated control qubits are $1$.
Here, the control qubit is in a superposition state, and it creates a quantum entanglement state, which plays a key role in GAS.
IQFT is applied at the last part of $\mathbf{A}_{y_i=0}$.
After that, the Grover operator $\mathbf{G}$ is applied $L$ times, and we measure the quantum state.
Fig.~\ref{fig:circuit-int-out} shows the probability that each state is measured, where the number of Grover operators was varied from $L=0$ to $2$.
The comma-separated text in this figure shows $n$ and $m$ qubits, and the latter is converted to a decimal number.
As shown in Fig.~\ref{fig:circuit-int-out}, when $L=0$, $2^3=8$ different states were observed with equal probability, and the corresponding values of the objective function were correctly calculated, demonstrating the potential of quantum computation.
When $L=1$ and $2$, only the state of interest $\b = [0~1~1]$, which yields $E(\b) = -1 < 0$, was successfully amplified by the Grover operator.
In this manner, GAS amplifies the states that are better than the current threshold and finds a binary solution that minimizes the objective function.

\subsection{Handling of Real-Valued Coefficients \cite{gilliam2021grover}\label{subsec:real}}
A polynomial may contain real-valued coefficients.
To deal with real-valued coefficients, Gilliam \etal proposed the following two methods \cite{gilliam2021grover}.

\subsubsection{Integer Approximation}
Multiplying the objective function by a positive constant does not affect the minimization process.
A real-valued coefficient can be approximated by multiplying a large number and rounding down to an integer.
Specifically, real coefficients are approximated as fractions with a common denominator, the denominator is multiplied to the objective function, and the numerators become approximated integer coefficients.
As can be inferred from \eqref{eq:mconstraint}, the drawback is that the number of required qubits $m$ increases as the value range of the objective function expands. 
If $m$ is kept small, this approximation becomes less accurate.

\subsubsection{Direct Encoding}
In this method, an integer $k$ in $\theta=2\pi k/2^m$ of \eqref{Ug-theta} is replaced with a real-valued coefficient $a \in \mathbb{R}$.
Then, the output probability indicates multiple integers, which is known as the Fej\'{e}r distribution.
Specifically, the state $\mathbf{U}_{\mathrm{Fej \acute{e}r}}(\theta)\Ket{0}_{m}$ after applying IQFT to $\mathbf{U}_G(\theta)\mathbf{H}^{\otimes m}\Ket{0}_{m}$ is given by \cite{gilliam2021grover}\footnote{This definition differs from \cite{gilliam2021grover}, but is essentially identical.}
\begin{align}
\label{U-Fejer}
\mathbf{U}_{\mathrm{Fej \acute{e} r}}(\theta)\Ket{0}_{m} = \sum_{l=0}^{2^m-1}\left\langle
\mathbf{g}(\theta),\mathbf{g}\left(2 \pi l  / 2^m \right)\right\rangle\Ket{l},
\end{align}
where we have $\theta = 2 \pi a / 2^m$ and $\mathbf{g}(\theta) = [1, e^{j\theta}, \cdots, e^{j(2^m-1)\theta}] / \sqrt{2^m}$.
The number of qubits $m$ must satisfy \cite{gilliam2021grover}
\begin{align}
-2^{m-1} \leq a < 2^{m-1}.\label{eq:mconstraintreal}
\end{align}
In this distribution, the probabilities of two integers close to a given real number $a$ are greater than the other probabilities.
For example, if $m=3$ qubits and $a=-2.5$, from \eqref{U-Fejer}, $-2$ and $-3$ are observed with equal probability.
If $a=-2.3$, $-2$ is observed more frequently than $-3$.

\subsection{Proposed GAS for Real-Valued HUBO}
As previously reviewed in Section~\ref{subsec:real}, in their innovative study \cite{gilliam2021grover}, Gilliam \etal proposed two methods for handling real-valued coefficients, but did not specifically investigate how GAS behaves in the case of direct encoding.
In such a case, in our evaluation, GAS samples a wrong value of the objective function, which obeys the Fej\'{e}r distribution.
For example, if the objective function value is $-2.5$, we may observe an integer value less than or equal to $-3$.
A value lower than the actual value is updated as the minimum and set as a new threshold $y_i$.
Then, no states satisfy $E(\b) - y_i < 0$, and one of all states is randomly sampled.
As a result, GAS will not be able to obtain an optimal solution.

\begin{algorithm}[tb]
    \caption{Proposed GAS designed for real-valued coefficients.\label{alg:real-gas}}
    \begin{algorithmic}[1]
        \renewcommand{\algorithmicrequire}{\textbf{Input:}}
        \renewcommand{\algorithmicensure}{\textbf{Output:}}
        \REQUIRE $E:\mathbb{B}^n\rightarrow\mathbb{R}, \lambda=8/7$
        \ENSURE  $\mathbf{b}$
        \STATE {Uniformly sample $\b_0 \in \mathbb{B}^n $ and set $y_0=E(\b_0)$}. \COMMENT{This step will be improved in Section~\ref{subsec:mvd}}
        \STATE {Set $k = 1$ and $i = 0$}.
        \REPEAT
        \STATE\hspace{\algorithmicindent}{Randomly select the rotation count $L_i$ from the set $\{0, 1, ..., \lceil k-1 \rceil$\}}.
        \STATE\hspace{\algorithmicindent}{Evaluate $\mathbf{G}^{L_i} \mathbf{A}_{y_i} \Ket{0}_{n+m}$, and obtain $\b$}. 
        \STATE\hspace{\algorithmicindent}{Evaluate $y=E(\b)$ in CD}. \COMMENT{This is the additional step}
        \hspace{\algorithmicindent}\IF{$y<y_i$}
        \STATE\hspace{\algorithmicindent}{$\b_{i+1}=\b, y_{i+1}=y,$ and $k=1$}.
        \hspace{\algorithmicindent}\ELSE{\STATE\hspace{\algorithmicindent}{$\b_{i+1}=\b_i, y_{i+1}=y_i,$ and $k=\min{\{\lambda k,\sqrt{2^n}}\}$}}.
        \ENDIF
        \STATE{$i=i+1$}.
        \UNTIL{a termination condition is met}.
    \end{algorithmic} 
\end{algorithm}
A possible solution here is that we ignore $y$ evaluated in QD.
Instead, we use $\b$ returned by GAS and calculate a correct objective function value $y = E(\b)$ in CD.
Since the quantum circuit using the direct encoding amplifies the states of interest with high probability, 
with this simple modification, GAS obtains an optimal solution correctly.
Overall, the above procedure is summarized in Algorithm~\ref{alg:real-gas}.

\begin{figure}[tb]
	\centering
    \includegraphics[clip, scale=0.58]{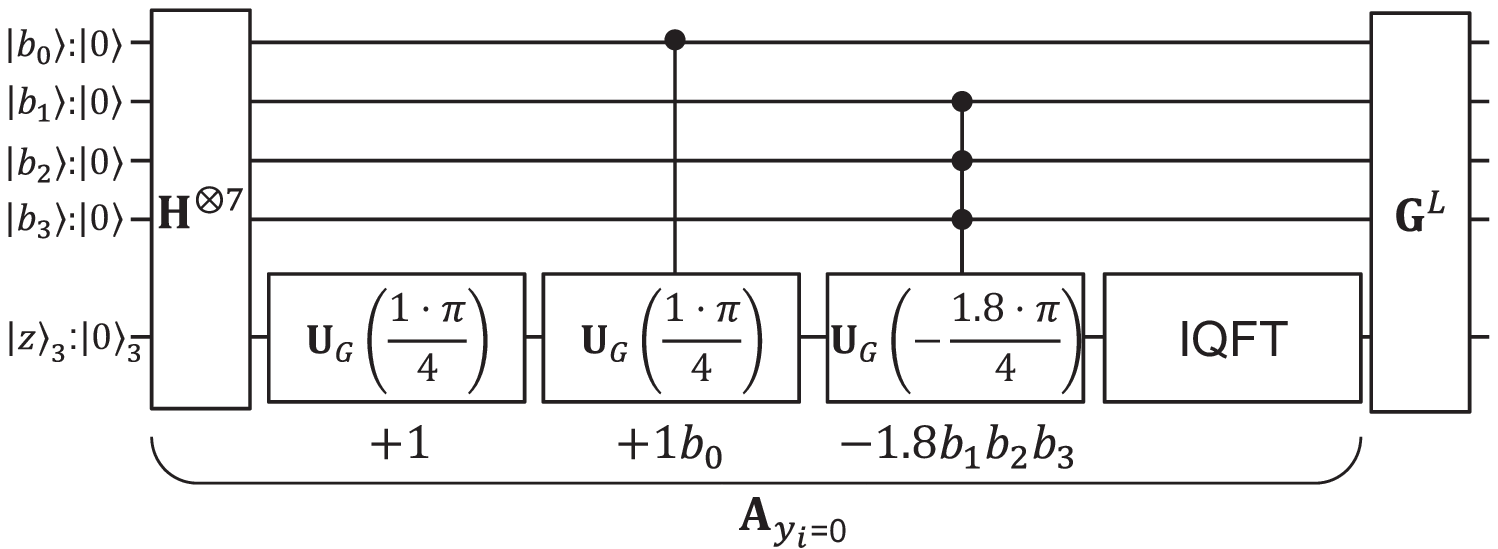}
	\caption{Quantum circuit corresponding to $E(\b)=1+b_0-1.8b_1b_2b_3$. \label{fig:circuit-real}}
	\subfigure[$L=0$.]{
		\includegraphics[clip, scale=0.65]{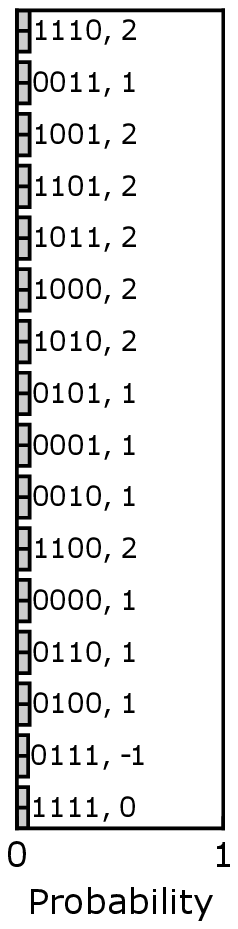}
	}
	\subfigure[$L=1$.]{
		\includegraphics[clip, scale=0.65]{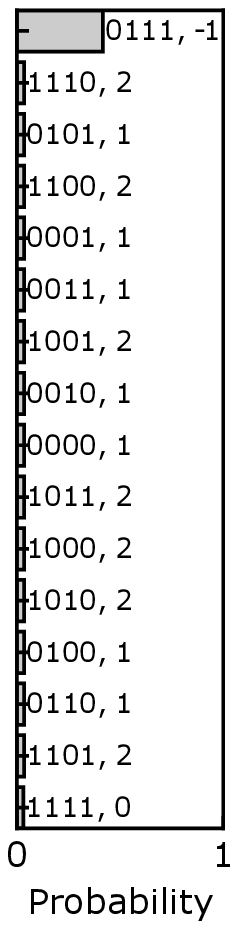}
	}
	\subfigure[$L=2$.]{
		\includegraphics[clip, scale=0.65]{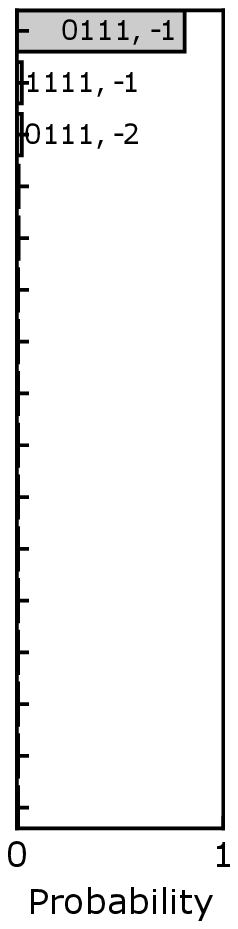}
	}
	\subfigure[$L=3$.]{
		\includegraphics[clip, scale=0.65]{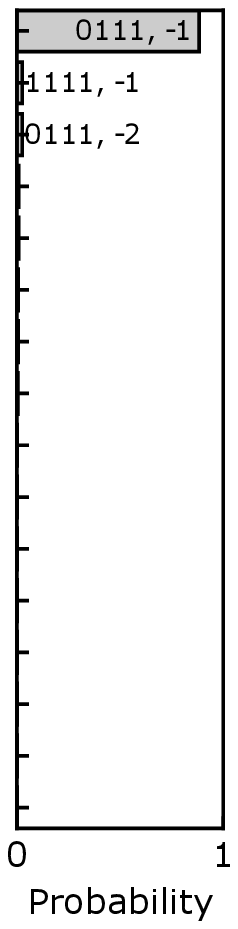}
	}
	\caption{Output probabilities of the circuit shown in Fig.~\ref{fig:circuit-real}, where only the top 16 states are shown.\label{fig:circuit-real-out}}
\end{figure}

We have two major drawbacks.
First, Algorithm~\ref{alg:real-gas} increases query complexity in CD, although the asymptotic order remains the same.
Second, the probability amplification may not be sufficient, which is illustrated in Fig.~\ref{fig:circuit-real-out}.

As a specific example, Fig.~\ref{fig:circuit-real} shows a quantum circuit corresponding to the objective function $E(\b)=1+b_0-1.8b_1b_2b_3$, where we set $n = 4$ and $m = 3$.
Since we used the direct encoding method, $-1.8 b_1b_2b_3$ was represented as $\U_G(-1.8 \pi / 4)$, and it was associated with three qubits $\Ket{b_1}$, $\Ket{b_2}$, and $\Ket{b_3}$.
Additionally, Fig.~\ref{fig:circuit-real-out} shows the output probabilities of Fig.~\ref{fig:circuit-real}, where only the top 16 states are shown for the sake of readability.
As given in \eqref{U-Fejer}, the direct encoding method may not result in a unique integer.
For example, the states $(0111,-1)$ and $(0111,-2)$ had positive probabilities in Fig.~\ref{fig:circuit-real}.
The state of interest here is $\b = [0 ~ 1 ~ 1 ~ 1]$ and $E(\b) = 1-1.8=-0.8 < 0$.
That is, before amplitude amplification, at $L = 0$, integers close to the real value are observed, and after amplification, at $L > 0$, the negative states are observed with high probabilities.
As shown in Fig.~\ref{fig:circuit-real-out}(d), the states
$(0111,-1)$ and $(0111,-2)$ were amplified as the number of Grover operators $L$ increased, while the wrong state $(0111,-2)$ was observed with a lower probability than $(0111,-1)$.
Another wrong state $(1111,-1)$ was also observed with a low probability.
This is the reason why the correction of the objective function value is required for real-valued GAS, as summarized in Algorithm~\ref{alg:real-gas}.

\subsection{Evaluation Metrics\label{subsec:metrics}}
In the literature, a quantum circuit has been evaluated by the numbers of qubits and gates and its depth, while a quantum algorithm has been evaluated by query complexity.

\subsubsection{Numbers of Qubits and Gates and Depth}
The size of the quantum circuit determines its feasibility.
As the numbers of qubits and gates in a quantum circuit increase, more advanced quantum computation becomes possible.
At the same time, however, it becomes more susceptible to noise and more difficult to implement in hardware.
In our evaluations, the number of required qubits is represented as $n+m$, and the number of quantum gates is derived as a function with respect to $n$ and $m$.

\subsubsection{Query Complexity \cite{botsinis2014fixedcomplexity}}
To investigate query complexity, we count how many times the objective function is queried.
Specifically, the query complexity in the classical domain (CD) is the number of times the objective function is evaluated, i.e., $i$ in Algorithm~\ref{alg:pro}.
By contrast, the query complexity in the quantum domain (QD) is the number of times the Grover operator $\G$ is applied, i.e., $L_0 + L_1 + \cdots + L_i$ in Algorithm~\ref{alg:pro}.
The definitions of query complexities in CD and QD are the same as those used in \cite{botsinis2014fixedcomplexity}.

\section{Quantum Speedup for MIMO MLD\label{sec:mld}}
Conventional studies on quantum-assisted wireless communications have not considered a specific construction method of the quantum circuit. In many cases, the circuit to calculate an objective function has been idealized as a black-box quantum oracle. In this section, we formulate the MIMO MLD as a new real-valued HUBO problem, which can be represented by a quantum circuit, as described in Section~\ref{sec:gas}.
We also analyze the probability distribution of the objective value for enabling further speedup.

\subsection{System Model\label{subsec:sys}}
\begin{figure}[tb]
	\centering
   \includegraphics[clip, scale=0.68]{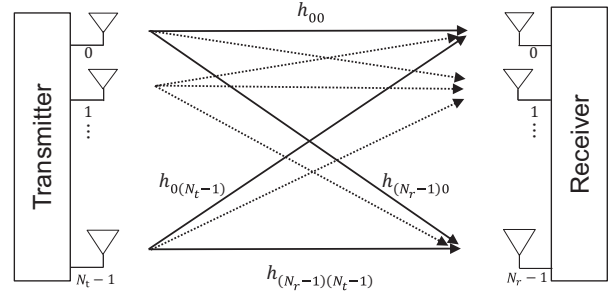}
	\caption{System model for MIMO with $N_{\mathrm{t}}$ transmit and $N_{\mathrm{r}}$ receiver antennas. \label{fig:mimo-model}}
\end{figure}
We consider a MIMO communication scenario with $N_{\mathrm{t}}$ transmit antennas and $N_{\mathrm{r}}$ receive antennas, as illustrated in Fig.~\ref{fig:mimo-model}.
The input $n$-bit sequence $\mathbf{b} = [b_0~b_1~\cdots~b_{n-1}] \in \mathbb{B}^n$ is mapped to a symbol vector $\mathbf{s} = [s_0~s_1~\cdots~s_{N_{\mathrm{t}}-1}] \in{\mathbb{C}}^{N_{\mathrm{t}}\times1}$, where $s_t$ for $0 \leq t \leq N_{\mathrm{t}}-1$ denotes a Gray-coded data symbol specified in 5G NR \cite{3gpp2018ts}.
We represent this bit-to-symbol mapper as $\mathbf{s} = M(\mathbf{b}) = M(b_0, \cdots, b_{n-1})$, which will be defined in detail in Section~\ref{subsec:mldhubo}.
The baseband received symbols $\mathbf{r} \in \mathbb{C}^{N_{\mathrm{r}} \times 1}$ is given by
\begin{align}
\mathbf{r} = \frac{1}{\sqrt{N_{\mathrm{t}}}} \mathbf{H}_{\mathrm{c}}\mathbf{s}+\sigma\mathbf{v},
\label{eq:r}
\end{align}
where $\mathbf{H}_{\mathrm{c}}\in{\mathbb{C}}^{N_{\mathrm{r}}\times N_{\mathrm{t}}}$ denotes the channel matrix, and $\mathbf{v}\in{\mathbb{C}}^{N_{\mathrm{r}}\times1}$ denotes the additive white Gaussian noise. 
Here, we assume the narrowband Rayleigh flat fading.
That is, each element of $\mathbf{H}_{\mathrm{c}}$, $h_{ut}$, and each element of $\mathbf{v}$, $v_u$, follow the standard complex Gaussian distribution $\mathcal{CN}(0, 1)$ for $0 \leq u \leq N_{\mathrm{r}} - 1$ and $0 \leq t \leq N_{\mathrm{t}} - 1$.
The SNR is defined as $\gamma = 1/\sigma^2$ because the symbol vector has the power constraint $\mathrm{E}\left[\| \mathbf{s} / \sqrt{N_{\mathrm{t}}} \|_{\mathrm{F}}^2 \right] = \mathrm{E}\left[\sum_{t=0}^{N_{\mathrm{t}}-1} |s_t|^2 / N_{\mathrm{t}} \right] = 1$.
The constellation size, or modulation order, is denoted by $L_{\mathrm{c}}$, and the transmission rate is calculated by
\begin{align}
    n = N_{\mathrm{t}} \log_2(L_{\mathrm{c}}) ~~ \text{[bit/symbol]}.
\end{align}

Corresponding to \eqref{eq:r}, the ideal MLD is performed as
\begin{align}
\hat{b}_0, \cdots, \hat{b}_{n-1} = \arg\underset{b_0, \cdots, b_{n-1}}{\min}
E(b_0, \cdots, b_{n-1}),
\label{eq:mld}
\end{align}
where we have the objective function
\begin{align}
E(b_0, \cdots, b_{n-1})=\left\|\mathbf{r}-\frac{1}{\sqrt{N_{\mathrm{t}}}}\mathbf{H}_{\mathrm{c}} M(b_0, \cdots, b_{n-1}) \right\|^{2}_{\mathrm{F}}.
\label{eq:obj}
\end{align}

From \eqref{eq:mld}, the exhaustive search using a classical computer requires the computational time complexity of $O(2^n)$, which is equivalent to the query complexity in CD.
Both complexities increase exponentially with the transmission rate $n$.

To mitigate the exponential complexity, a number of low-complexity detectors have been proposed in the literature.
The classic ZF detector uses the pseudo-inverse matrix of 
\begin{align}
\mathbf{W}_{\mathrm{ZF}}=
\begin{cases}(\mathbf{H}_{\mathrm{c}}^{\mathrm{H}}\mathbf{H}_{\mathrm{c}})^{-1}\mathbf{H}_{\mathrm{c}}^{\mathrm{H}}& (N_{\mathrm{t}}\leq N_{\mathrm{r}})\\
\mathbf{H}_{\mathrm{c}}^{\mathrm{H}}(\mathbf{H}_{\mathrm{c}}\mathbf{H}_{\mathrm{c}}^{\mathrm{H}})^{-1} & (N_{\mathrm{t}}>N_{\mathrm{r}})
\end{cases}
\end{align}
and enables independent detection of data symbols as
\begin{align}
\hat{b}_0, \cdots, \hat{b}_{n-1} = M^{-1} (\mathbf{W}_{\mathrm{ZF}}\mathbf{r}),
\end{align}
where $M^{-1} \left( \cdot \right)$ denotes the hard-decision symbol-to-bit demapper.
Similarly, the MMSE detector uses
\begin{align}
\mathbf{W}_{\mathrm{MMSE}}=
\begin{cases}(\mathbf{H}_{\mathrm{c}}^{\mathrm{H}}\mathbf{H}_{\mathrm{c}}+\sigma^2\I)^{-1}\mathbf{H}_{\mathrm{c}}^{\mathrm{H}}& (N_{\mathrm{t}}\leq N_{\mathrm{r}})\\
\mathbf{H}_{\mathrm{c}}^{\mathrm{H}}(\mathbf{H}_{\mathrm{c}}\mathbf{H}_{\mathrm{c}}^{\mathrm{H}}+\sigma^2\I)^{-1} & (N_{\mathrm{t}}>N_{\mathrm{r}})
\label{eq:MMSE}
\end{cases}
\end{align}
and obtains
\begin{align}
\hat{b}_0, \cdots, \hat{b}_{n-1} = M^{-1} (\mathbf{W}_{\mathrm{MMSE}}\mathbf{r}).
\end{align}
An MMSE-based interference cancelation method has been adopted in typical wireless standards such as 5G NR.
The performance of a ZF or MMSE detector is worse than that of MLD.
In general, low-complexity detectors improve complexity at the sacrifice of performance.

The above system model and detectors are typical and common in the field of wireless communications.
Since we consider a general MIMO system, the simulation results given in this paper are the same as those for a multicarrier scenario without inter-subcarrier interference or an uplink multi-user scenario in which $N_{\mathrm{t}}$ single-antenna user terminals transmit their symbols and these symbols are received simultaneously at a base station equipped with $N_{\mathrm{r}}$ antennas.

\subsection{Proposed Method to Transform MLD into HUBO\label{subsec:mldhubo}}
As described in Section~\ref{sec:gas}, the proposed GAS is capable of solving a real-valued HUBO problem.
We transform the objective function of MIMO MLD \eqref{eq:obj} into a HUBO problem.
Specifically, we use the relationship between transmission bits and data symbols, which is specified in the 5G NR standard \cite{3gpp2018ts}.
The input $n$-bit sequence is denoted by $\mathbf{b} = [b_0 ~ b_1 ~ \cdots ~ b_{n-1}] \in \mathbb{B}^n$ and 
the symbol vector is denoted by $\mathbf{s} = [s_0 ~ s_1 ~ \cdots ~ s_{N_{\mathrm{t}} - 1}]\in \mathbb{C}^{N_{\mathrm{t}}}$.
Then, BPSK symbols $\mathbf{s} = M_2(\mathbf{b})$ are generated by \cite{3gpp2018ts}
\begin{align}
\label{map-bpsk}
s_t = \frac{1}{\sqrt2}[(1-2b_t)+j(1-2b_t)]
\end{align}
and QPSK symbols $\mathbf{s} = M_4(\mathbf{b})$ are generated by \cite{3gpp2018ts}
\begin{align}
\label{map-qpsk}
s_t = \frac{1}{\sqrt2}[(1-2b_{2t})+j(1-2b_{2t+1})].
\end{align}
Furthermore, 16-QAM symbols $\mathbf{s} = M_{16}(\mathbf{b})$ are generated by \cite{3gpp2018ts}
\begin{align}
\label{map-16qam}
s_t=&\frac{1}{\sqrt{10}}(1-2b_{4t+0})[2-(1-2b_{4t+2})] \notag \\ 
+&\frac{j}{\sqrt{10}}(1-2b_{4t+1})[2-(1-2b_{4t+3})]
\end{align}
and 64-QAM symbols $\mathbf{s} = M_{64}(\mathbf{b})$ are generated by \cite{3gpp2018ts}
\begin{align}
\label{map-64qam}
s_t=&\frac{1}{\sqrt{42}}(1-2b_{6t+0})[4-(1-2b_{6t+2})[2-(1-2b_{6t+4})]] \notag \\ 
+&\frac{j}{\sqrt{42}}(1-2b_{6t+1})[4-(1-2b_{6t+3})[2-(1-2b_{6t+5})]].
\end{align}
A similar relationship for 256-QAM is defined in \cite{3gpp2018ts} and its extension for higher modulation orders can be defined easily.
\begin{figure}[tb]
	\centering
   \includegraphics[clip, scale=0.68]{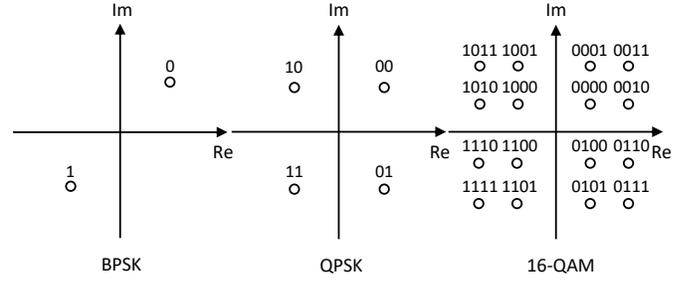}
	\caption{Constellation for Gray-coded data symbols specified in the 5G NR standard \cite{3gpp2018ts} \label{fig:map-mod}}
\end{figure}
Overall, Fig.~\ref{fig:map-mod} shows the Gray-coded data symbols defined by \eqref{map-bpsk}, \eqref{map-qpsk}, and \eqref{map-16qam}.

Our proposed objective function is obtained by substituting $M(\cdot)$ in \eqref{eq:obj} with \eqref{map-bpsk}, \eqref{map-qpsk}, \eqref{map-16qam}, or \eqref{map-64qam}, which contains $n$ number of binary variables $b_0, \cdots, b_{n-1}$.
In the cases of BPSK and QPSK, our objective function results in a quadratic form since both symbols are represented by a linear relationship and the MLD \eqref{eq:mld} contains the square of the Frobenius norm.
In the case of 16-QAM, the objective function results in a quartic form since the symbols are represented by a quadratic relationship.
Similarly, the objective function results in a sextic form in the 64-QAM case.

The use of data symbols specified in 5G NR is not straightforward since the objective function inevitably contains higher-order terms if the modulation order is 16 or higher.
Thus, in this form, the conventional QA requires a transformation from HUBO to QUBO, and this transformation involves an increase in binary variables, making the problem more difficult.
Our proposed approach is only possible with the aid of the real-valued support of GAS.
Because of GAS, the query complexity is expected to be reduced from $O(2^n)$ to $O(\sqrt{2^n})$.

The structure of the proposed objective function depends only on the number of transmit antennas, $N_{\mathrm{t}}$, the number of receive antennas, $N_{\mathrm{r}}$, and the modulation order, $L_{\mathrm{c}}$.
The coefficients in the objective function change depending on the channel matrix $\mathbf{H}_{\mathrm{c}}$.
The calculation cost of coefficients determines the complexity of classical processing required before executing GAS, which relates to the latency of the algorithm.
If we approximate the computational complexity as the number of real-valued multiplications, the largest burden is the product of channel coefficients, such as $h_{00}h_{01}^*$.
The total number of multiplications is calculated as $4N_{\mathrm{r}} N_{\mathrm{t}}(N_{\mathrm{t}} - 1)/2 = O(N_{\mathrm{r}} N_{\mathrm{t}}^2)$, which is sufficiently small with respect to the detection complexity.

\paragraph*{Example (QPSK)}
As a specific example, we consider the QPSK case \eqref{map-qpsk} with $N_{\mathrm{t}} = N_{\mathrm{r}} = 2$. The objective function of \eqref{eq:obj} can be transformed into
\begin{align}
\label{qpsk-obj}
&E(b_0, b_1, b_2, b_3) \notag \\
=& 2\sum_{u=0}^{1} {\sum_{t=0}^{1}({\mathrm{Re}(h_{ut}r_u^*)}b_{2t}-{\mathrm{Im}(h_{ut}r_u^*)}b_{2t+1}}) \notag \\
+& 2a_1(b_0 b_2+b_1 b_3)+2a_2(b_0 b_3-b_1 b_2) \notag \\ 
-& (a_1+a_2)(b_0+b_3)-(a_1-a_2)(b_1+b_2),
\end{align}
where we have $a_1=\mathrm{Re}(h_{00}h_{01}^*)+\mathrm{Re}(h_{10}h_{11}^*)$ and $a_2=\mathrm{Im}(h_{00}h_{01}^*)+\mathrm{Im}(h_{10}h_{11}^*)$.
This function \eqref{qpsk-obj} is in a quadratic form.

\paragraph*{Example (16-QAM)}
\begin{figure*}[tb]
	\centering
    \includegraphics[clip, scale=0.175]{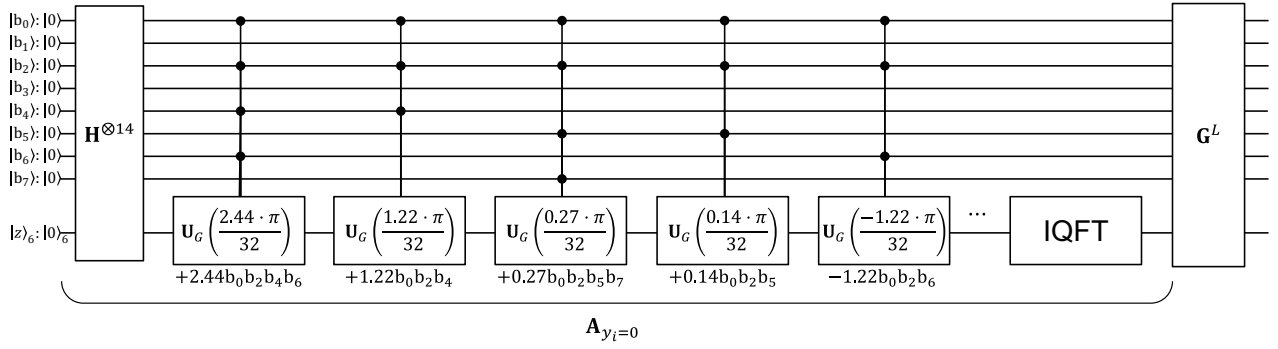}
	\caption{Quantum circuit corresponding to objective function of 16-QAM detection. \label{fig:circuit-16qam}}
\end{figure*}
Additionally, Fig.~\ref{fig:circuit-16qam} exemplifies a specific quantum circuit for the 16-QAM case with $N_{\mathrm{t}} = N_{\mathrm{r}} = 2$, where we have $n=8$ qubits for binary variables, $m=5$ qubits for real-valued encoding, random channel coefficients
\begin{align}
    \mathbf{H}_{\mathrm{c}} = [[&0.748510757437062 - 0.014877263039446401j, \nonumber \\
    &1.3215983896521515 + 0.06298233870206783j], \nonumber \\
    [&0.6371630706424066 - 0.14262155021296025j, \nonumber \\
    -&0.3888005272494009 - 0.15170387681055802j]],
    \label{eq:exHc}
\end{align}
and the original information bits of $00110101$.
As shown in Fig.~\ref{fig:circuit-16qam}, the objective function results in a quartic form: $E(\b)=1.22b_0b_2b_4b_6+0.61b_0b_2b_4+\cdots$ using \eqref{eq:obj} and \eqref{map-16qam}.
As an example, the coefficient of $b_0b_2b_4b_6$ is calculated as $\frac{1}{2}\cdot\frac{4}{\sqrt{10}}\cdot \frac{4}{\sqrt{10}}(h_{00}h_{01}^* +h_{00}^* h_{01}+h_{10}h_{11}^* +h_{10}^* h_{11}) = 1.22$, which is rounded down to the second decimal place for simple illustration.

\subsection{Proposed Threshold for Further Speedup\label{subsec:mvd}}
GAS obtains a global minimum solution by updating the threshold value $y_i$ and amplifying the probability amplitudes corresponding to values smaller than the threshold.
The query complexity can be reduced by setting the initial threshold in a manner different from that in classic random sampling, although the asymptotic performance may not change.
In this section, we derive the probability distribution of the objective function value and use it to determine a strict threshold, which enables further speedup.

If the information bits in \eqref{eq:obj} are estimated correctly, the minimum value of \eqref{eq:obj} is the Frobenius norm of additive noise $\mathbf{v} \in \mathbb{C}^{N_{\mathrm{r}} \times 1}$ as follows:
\begin{align}
\label{min-obj}
E_{\mathrm{min}}=
\underbrace{\sigma^2}_{\text{known}}
\underbrace{\sum_{u=0}^{N_{\mathrm{r}}-1}|v_u|^2}_{\text{unknown}}.
\end{align}
That is, $E_{\mathrm{min}}$ depends on the noise variance $\sigma^2$, which is typically known at the receiver, and instantaneous noise $v_u$, which is unknown in any case.
Since the noise is assumed to follow the complex Gaussian distribution, the magnitude of the norm follows the Rayleigh distribution, and its square follows the exponential distribution.
%As a result, the probability density function of $E_{\mathrm{min}}$ in \eqref{min-obj} is the Erlang distribution
As a result, $E_{\mathrm{min}}$ in \eqref{min-obj} follows the Erlang distribution, whose probability density function is
\begin{align}
f(y)=\frac{\gamma^{N_{\mathrm{r}}} y^{N_{\mathrm{r}}-1} e^{-\gamma y}}{(N_{\mathrm{r}}-1)!},
\end{align}
where we have SNR $\gamma=1/\sigma^2$.
The corresponding cumulative distribution function (CDF) is given by
\begin{align}
\label{cdf-e}
F(y)=\mathrm{Pr}[Y\leq y]=1-e^{-\gamma y}\sum_{u=0}^{N_{\mathrm{r}}-1}\frac{(\gamma y)^u}{u!}.
\end{align}

\begin{figure}[tb]
	\centering
   \includegraphics[clip, scale=0.6]{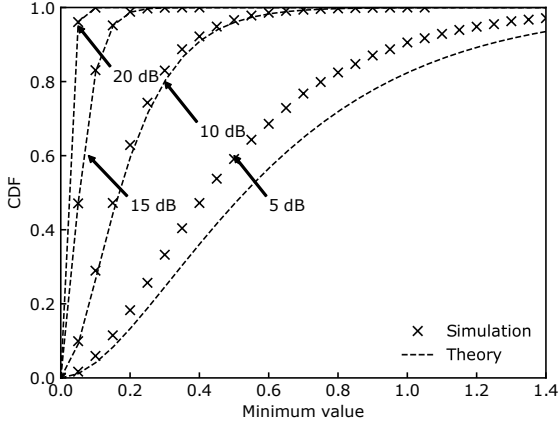}
	\caption{Cumulative distribution of the minimum of objective function values \eqref{eq:cdf-e-2}.\label{fig:cdf_min}}
\end{figure}
As an example, if we consider the case with $N_{\mathrm{r}}=2$, the CDF is calculated as
\begin{align}
F(y)=\mathrm{Pr}[Y\leq y]=1-e^{-\gamma y}(1+\gamma y)
\label{eq:cdf-e-2}
\end{align}
from \eqref{cdf-e}.
Fig.~\ref{fig:cdf_min} exemplifies CDF \eqref{eq:cdf-e-2} when SNR is varied as $\gamma = 5$ to $20$ dB.
Additionally, the CDF of the simulated objective function values with $N_{\mathrm{t}} = 2$ and QPSK is also plotted.
As shown in Fig.~\ref{fig:cdf_min}, if the SNR is sufficient, such as above 10 dB, the theoretical and simulated values are identical.
Thus, it is possible to know in advance that the minimum value to be calculated is below a certain threshold, which can be determined with a very high degree of certainty.
Given an SNR $\gamma$, theoretical values of \eqref{cdf-e} can be used to determine a strict threshold.

From \eqref{cdf-e}, the probability that the threshold $y$ is below the minimum value is
\begin{align}
\mathrm{Pr}[Y>y]=e^{-\gamma y}(1+\gamma y).\label{eq:PYy}
\end{align}
Let $\tilde{y}$ be the threshold to be determined and $P$ be a small constant probability, such as $P=10^{-3}$ and $10^{-4}$.
Replacing $y$ and $\mathrm{Pr}[Y>y]$ in \eqref{eq:PYy} with $\tilde{y}$ and $P$ yields
\begin{align}
P=e^{-\gamma \tilde{y}}(1+\gamma \tilde{y}).
\end{align}
Dividing both sides by $-e$ gives
\begin{align}
-\frac{P}{e}=-(1+\gamma \tilde{y})e^{-(1+\gamma \tilde{y})}.
\end{align}
Then, using the Lambert $W$ function, we obtain
\begin{align}
W_{-1}\left(-\frac{P}{e}\right)=-(1+\gamma \tilde{y}) = - \left(1+ \frac{\tilde{y}}{\sigma^2} \right),
\end{align}
where $W_{-1}(\cdot)$ denotes the lower branch of the Lambert $W$ function, i.e., $W_{-1}(\cdot) \leq -1$ and $W_{-1}(-1/e) = -1$.
Finally, the threshold to be determined is
\begin{align}
\tilde{y} = 
\underbrace{\sigma^2}_{\text{known}}
\underbrace{\nu}_{\text{known}},
\label{eq:ym}
\end{align}
which is similar to \eqref{min-obj},
and 
\begin{align}
\nu = -1 - W_{-1}\left(-\frac{P}{e}\right).
\end{align}
Here, $\nu$ is a positive constant and is calculated once before running our proposed algorithm.
For example, we have $\nu = 9.23$ if $P = 10^{-3}$ and $\nu = 11.8$ if $P = 10^{-4}$.

For further speedup, we opt to use the output of the MMSE detector \eqref{eq:MMSE}.
In \cite{botsinis2014fixedcomplexity}, Botsinis \etal proposed the MMSE-based threshold of
\begin{align}
    \bar{y} = E(\bar{\mathbf{b}}_0),
\end{align}
where we have a rough estimate $\bar{\mathbf{b}}_0 = M^{-1}(\mathbf{W}_{\mathrm{MMSE}} \mathbf{r})$.
Our proposed threshold $\tilde{y}$, which is simpler than $\bar{y}$, can be used together with $\bar{y}$.
Specifically, we calculate both $\tilde{y}$ and $\bar{y}$ at the beginning of Algorithm~\ref{alg:real-gas}, set the initial threshold as the smaller of the two, and initialize the first solution with $\bar{\mathbf{b}}_0$.
Let $\mathbf{b}_0$ be a random $n$-bit sequence.
The initial threshold used for the proposed Algorithm~\ref{alg:real-gas} can be summarized as follows:
\begin{align}
    y_0 = \begin{cases}
        E(\mathbf{b}_0) & \text{(Original GAS \cite{gilliam2021grover})}\\
        \bar{y} & \text{(MMSE-based threshold \cite{botsinis2014fixedcomplexity})}\\
        \tilde{y} & \text{(Proposed threshold)}\\
        \min(\bar{y}, \tilde{y}) & \text{(Proposed combination)}\\
    \end{cases}.
\end{align}

One problem with the proposed threshold $\tilde{y}$ and the combination $\min(\bar{y}, \tilde{y})$ is that they may become smaller than the actual minimum.
In this case, since there are no states of interest, GAS will be in a state where the solution $\mathbf{b}_i$ in Algorithm~\ref{alg:real-gas} is not updated.
The probability of this undesirable event occurring is $P$, i.e.,
\begin{align}
	\mathrm{Pr}[\tilde{y} < E_{\mathrm{min}}] = \mathrm{Pr}[\min(\bar{y}, \tilde{y}) < E_{\mathrm{min}}] = P,
\end{align}
because we have the relationship $\mathrm{Pr}[\bar{y} < E_{\mathrm{min}}] = 0$.
Then, it can be expected that the proposed threshold $\tilde{y}$ may degrade the bit error ratio (BER) significantly if $P$ is inappropriate.
Specifically, the BER of the proposed threshold $\tilde{y}$ is approximated by
\begin{align}
	P \cdot 0.5 + (1-P) \cdot \mathrm{BER}_{\mathrm{MLD}},\label{eq:BERprop}
\end{align}
where $\mathrm{BER}_{\mathrm{MLD}}$ is the BER of MLD.
In the proposed combination method, we initialize the first solution with the MMSE output $\bar{\mathbf{b}}_0$.
Since the initial threshold $\min(\bar{y}, \tilde{y})$ becomes smaller than the actual minimum with probability $P$, the BER of the proposed combination method is approximated by
\begin{align}
	P \cdot \mathrm{BER}_{\mathrm{\mathrm{MMSE}}} + (1-P) \cdot \mathrm{BER}_{\mathrm{MLD}},\label{eq:BERcomb}
\end{align}
where $\mathrm{BER}_{\mathrm{\mathrm{MMSE}}}$ is the BER of the MMSE detector.
Both \eqref{eq:BERprop} and \eqref{eq:BERcomb} indicate that 
the design of $P$ exerts no significant effect as long as it is smaller than $\mathrm{BER}_{\mathrm{MLD}}$, which can be calculated exactly in a closed form in advance.
For our performance analysis, the effect of $P$ is shown in Fig.~\ref{fig:iniber-20dB}.

\section{Performance Analysis\label{sec:comp}}
In this section, we analyze the number of quantum gates required by GAS, which is represented as a function of the numbers of qubits $n$ and $m$.
Then, we investigate the performance of the proposed formulation in terms of BER and evaluate the proposed algorithm in terms of the rate of convergence.
Here, both integer approximation and direct encoding are considered.
Finally, we evaluate the effects of the proposed threshold.

\subsection{Algebraic Analysis of the Number of Quantum Gates\label{subsec:ngates}}
A quantum circuit for GAS is composed of $\mathbf{H}$, $\mathbf{X}$, $\mathbf{Z}$, phase, controlled-phase gates, and the IQFT.
In particular, the state preparation operator $\mathbf{A}_{y_i}$ is the most complex part corresponding to the objective function and is dynamically configured in accordance with the threshold $y_i$.
In the quantum circuit $\mathbf{A}_{y_i}$, the number of controlled-phase gates depends on the number of terms in the objective function.
We therefore derive the number of terms in the objective function that correspond to each order in an algebraic manner.
Ignoring the power scaling factor,  the objective function of MIMO MLD \eqref{eq:obj} is transformed into
\begin{align}
&\sum_{u=0}^{N_{\mathrm{r}}-1}|r_u-h_{u0}s_0-h_{u1}s_1-\cdots-h_{u(N_{\mathrm{t}}-1)}s_{N_{\mathrm{t}}-1}|^2\notag\\
=&\sum_{u=0}^{N_{\mathrm{r}}-1}(r_u-h_{u0}s_0-h_{u1}s_1-\cdots-h_{u(N_{\mathrm{t}}-1)}s_{N_{\mathrm{t}}-1})\notag\\
&\cdot {(r_u-h_{u0}s_0-h_{u1}s_1-\cdots-h_{u(N_{\mathrm{t}}-1)}s_{N_{\mathrm{t}}-1})}^*.
\end{align}
Here, we focus on three types of terms: first-order terms such as $-r^*_0h_{00}s_0$ and $-r_0h^*_{00}s^*_0$, 
squares of the same symbol such as $|h_{00}|^2|s_0|^2$ and $|h_{01}|^2|s_1|^2$, and products of two symbols such as $h_{00}h^*_{10}s_0s^*_1$ and $h^*_{00}h_{10}s^*_0{s}_1$.

For example, in the relatively simple QPSK case, squares of the same symbol result in constant terms because of \eqref{map-qpsk}. First-order terms directly result in first-order terms with respect to binary variables. Products between two symbols result in products of binary variables.
If $N_{\mathrm{t}} = 2$ and $N_{\mathrm{r}} = 2$, four second-order terms appear: $b_0b_2, b_1b_3, b_0b_3,$ and $b_1b_2$.
The number of corresponding terms is equal to the combination of two choices from $N_{\mathrm{t}}$ antennas, e.g., 
\begin{align}
{N_{\mathrm{t}} \choose 2} = \frac{N_{\mathrm{t}}(N_{\mathrm{t}}-1)}{2}=\frac{n(n-2)}{8},
\end{align}
where we have the relationship $n = N_{\mathrm{t}} \cdot \log_2(L_{\mathrm{c}}) = 2N_{\mathrm{t}}$.
In total, the number of second-order terms is calculated as $4 \cdot n(n-2) / 8 = n(n-2)/2$.

\begin{table*}[tb]
	\centering
	\caption{Number of quantum gates required for $\mathbf{A}_{y_i}$ ($n$-bit transmission with $m$-bit accuracy) \label{table:count-gate}}
	\scalebox{0.85}[0.8]{ % page reduction
	\begin{tabular}{l ll ll ll ll}
	    \hline
        Gate & BPSK&& QPSK&& 16-QAM&& 64-QAM& \\
		\hline
		H   & $n+m$&$=O(n+m)$ & $n+m$&$=O(n+m)$ &$n+m$&$=O(n+m)$&$n+m$&$=O(n+m)$ \\
	    R   & $m$&$=O(m)$     & $m$&$=O(m)$     &$m$&$=O(m)$&$m$&$=O(m)$\\
	    1-CR&$nm$&$=O(nm)$    & $nm$&$=O(nm)$   &$nm$&$=O(nm)$&$nm$&$=O(nm)$\\
	    2-CR&${n(n-1)m}/2$&$=O(n^2m)$ & ${n(n-2)m}/2$&$=O(n^2m)$&${n(n-3)m}/2$&$=O(n^2m)$&${n(n-4)m}/2$&$=O(n^2m)$ \\
	    3-CR&$0$&             &$0$&&${n(n-4)m}/2$&$=O(n^2m)$&$n(n-6)m+nm/3$&$=O(n^2m)$ \\
	    4-CR&$0$&             &$0$&&$n(n-4)m/8$&$=O(n^2m)$&$5n(n-6)m/6$&$=O(n^2m)$ \\
	    5-CR&$0$&             &$0$&&$0$&&$n(n-6)m/3$&$=O(n^2m)$ \\
	    6-CR&$0$&             &$0$&&$0$&&$n(n-6)m/18$&$=O(n^2m)$ \\
	    IQFT&$1$&             &$1$&&$1$&&$1$& \\
		\hline
	\end{tabular}
	}
\end{table*}
Extending the QPSK case, we counted the number of terms in the objective function for each modulation order and derived the number of quantum gates required by GAS.
Table~\ref{table:count-gate} summarizes the derived results, where the quantum gates were categorized by type.
As given in Table~\ref{table:count-gate},
the number of controlled-phase gates mainly depends on the number of binary variables $n$.
Here, 1-CR represents the controlled-phase gate, and 2-CR, 3-CR, $\cdots$ represent multi-controlled-phase gates.
Since we have the relationship $n = N_{\mathrm{t}} \cdot \log_2(L_{\mathrm{c}})$, 
the quantum circuit becomes increasingly complex depending on the square of the number of antennas $N_{\mathrm{t}}$ and the modulation order $L_{\mathrm{c}}$.

We analyze the number of quantum gates in the entire circuit $\mathbf{G}^{L_i} \mathbf{A}_{y _i}\Ket{0}_{n+m}$, where we have $\mathbf{G}=\mathbf{A}_{y_i}\mathbf{D}\mathbf{A}_{y_i}^{\mathrm{H}} \mathbf{O}$.
In each iteration, the Grover operator is applied $L_i$ times, where $L_i$ is a uniform random number.
$\mathbf{O}$ is composed of a single $\mathbf{Z}$ gate and $\mathbf{D}$ is the Grover diffusion operator, each of which is repeated $L_i$ times.
The other part contains $(2L_i+1)(n+m)$ $\mathbf{H}$ gates, $(2L_i+1)m$ phase gates, $(2L_i+1)c$ controlled-phase gates, and $(2L_i+1)$ IQFT, where $c$ is the number of controlled-phase gates given in Table~\ref{table:count-gate}.
In summary, $N_{\mathrm{t}}$ and $L_{\mathrm{c}}$ affect the number of gates by the power of two, while $m$ and $L_i$ affect it in direct proportion. 

Real quantum computers rely on decomposed elementary gates: single unitary gates and controlled NOT gates \cite{barenco1995elementary}.
Specifically, a phase gate with $c$ control qubits is decomposed into $O(c)$ elementary gates.
Thus, according to Table~\ref{table:count-gate}, the number of elementary gates required for controlled-phase gates is $O(cn^2m)$ in total.
Additionally, IQFT requires $O(m^2)$ elementary gates \cite{nielsen2010quantum}.
Here, the only parameters that can be designed are $n$ and $m$.
With the aid of our efficient HUBO formulation, the number of qubits $n$ cannot be further reduced, since the search space size of MIMO ML detection is $2^n$.
Later, we investigate whether our real-valued GAS can reduce $m$.

\subsection{Effects of Integer Approximation\label{subsec:int}}
\begin{figure}[tb]
	\centering
    \includegraphics[clip, scale=0.6]{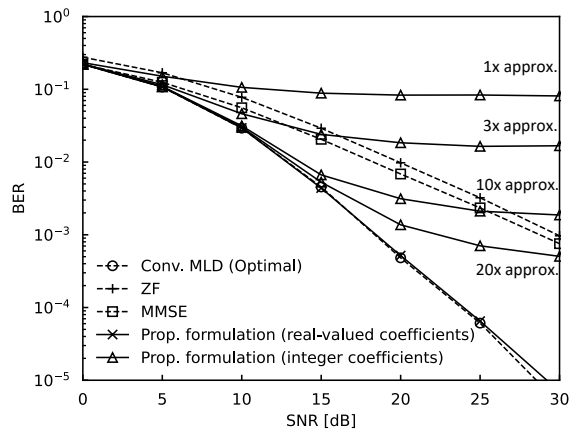}
	\caption{BER for the QPSK case with $N_{\mathrm{t}}=N_{\mathrm{r}}=2$.\label{fig:qpsk-ber}}
\end{figure}
First, Fig.~\ref{fig:qpsk-ber} shows BER of the classic MLD and the proposed formulations that consider the integer approximation with different accuracies.
Specifically, the real values were multiplied by 1, 3, 10 or 20, and approximated by rounding them to the nearest integers.
As references, the BER curves of ZF, MMSE, and the real-valued formulation were also plotted.
To analyze the effects of approximation accuracy, BER values were calculated using the state-of-the-art optimization solver, IBM CPLEX, instead of quantum simulations.
As shown in Fig.~\ref{fig:qpsk-ber}, BER performance varied significantly depending on the accuracy of the conventional integer approximation.
High approximation accuracy leads to large integers, resulting in an increase in the number of qubits $m$.
In contrast, the proposed real-valued formulation achieved the same performance as the classic MLD.
This observation indicates that the proposed real-valued GAS algorithm must be invoked to solve the MIMO MLD problem on a quantum computer.

\begin{figure}[tb]
	\centering
	\subfigure[QPSK ($3$x approximation, $n=4$, $m=6$ qubits).]{
		\includegraphics[clip, scale=0.6]{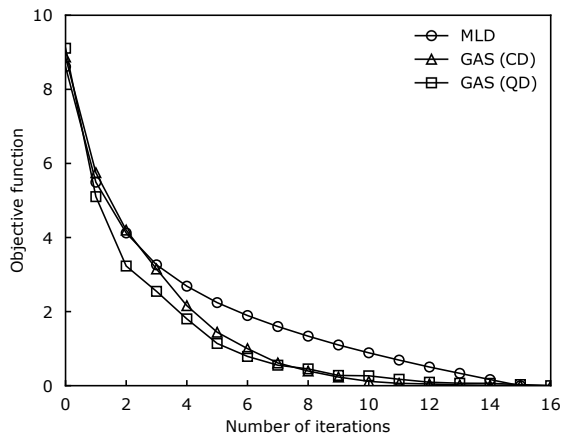}%page reduction
	}
	\subfigure[16-QAM ($14$x approximation, $n=8$, $m=8$ qubits).]{
		\includegraphics[clip, scale=0.6]{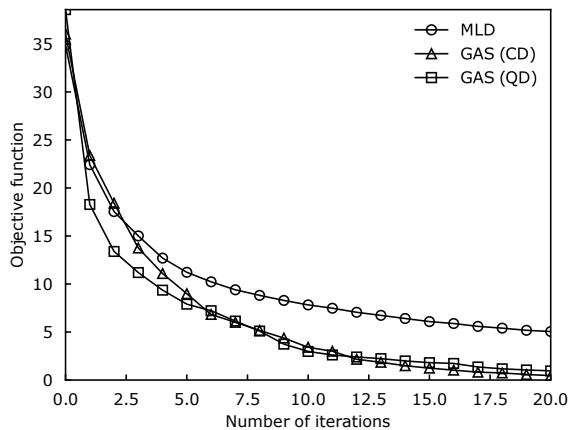}%page reduction
	}
	\caption{Average objective function values with the integer approximation and $N_{\mathrm{t}}=N_{\mathrm{r}}=2$. \label{fig:int-obj}}
\end{figure}
Next, Fig.~\ref{fig:int-obj} shows the average objective function values when increasing the number of iterations, where iterations in both CD and QD were considered.
We assumed a sufficiently high SNR and the fixed $2 \times 2$ channel matrix given in \eqref{eq:exHc}.\footnote{Note that we observed the same trend for different channel coefficients and SNRs.}
We used the original GAS with a random initial threshold and terminated the simulation if the objective function value remained the same more than 20 times in CD.
In Fig.~\ref{fig:int-obj}(a), real values were multiplied by 3 and rounded down to integers, and in Fig.~\ref{fig:int-obj}(b), real values were multiplied by 14 and were approximated.
The number of qubits $m$ required for encoding the value $E(\b) - y_i$ was set to an integer sufficient not to overflow, i.e., $m=6$ in Fig.~\ref{fig:int-obj}(a) and $m=8$ in Fig.~\ref{fig:int-obj}(b).
Note again that the integer approximation requires more qubits to encode the value.
Because quantum simulations with $n+m=16$ qubits are time-consuming, we fixed the input bits to $00110101$ in Fig.~\ref{fig:int-obj}(b), while the bits were generated randomly in Fig.~\ref{fig:int-obj}(a).
For a clear illustration, we added a constant value to the objective function so that $E_{\mathrm{min}} = 0$.
It was observed in Fig.~\ref{fig:int-obj}(a) that the query complexities of GAS in CD and QD were almost the same as in the exhaustive search of MLD.
By contrast, in Fig.~\ref{fig:int-obj}(b), GAS exhibited better query complexities in both CD and QD than did MLD.
That is, the quantum advantage improved as the problem size increased.

\subsection{Effects of Direct Encoding\label{subsec:effreal}}
\begin{figure}[tb]
	\centering
	\subfigure[QPSK ($n=4$, $m=5$ qubits).]{
		\includegraphics[clip, scale=0.6]{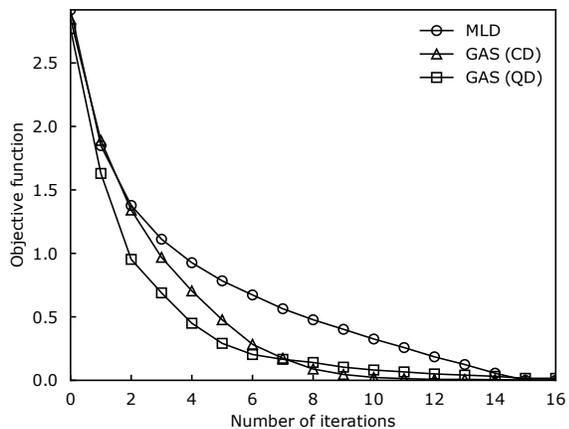}%page reduction
	}
	\subfigure[16-QAM ($n=8$, $m=5$ qubits).]{
		\includegraphics[clip, scale=0.6]{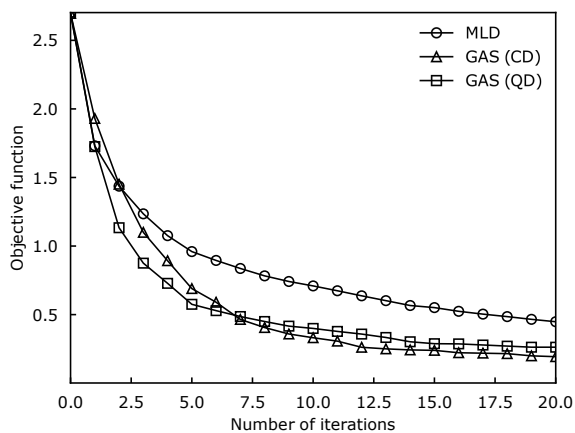}%page reduction
	}
	\caption{Average objective function values with direct encoding and $N_{\mathrm{t}}=N_{\mathrm{r}}=2$. \label{fig:real-obj}}
\end{figure}
Similar to Fig.~\ref{fig:int-obj}, Fig.~\ref{fig:real-obj} shows the average objective function values when increasing the number of iterations in CD and QD, where we used direct encoding.
The simulation parameters were the same as those used in Fig.~\ref{fig:int-obj} except for the real-valued expression and the number of required qubits $m$.
Specifically, the number of qubits $m=6$ in Fig.~\ref{fig:int-obj}(a) was reduced to $m=5$ in Fig.~\ref{fig:real-obj}(a).
Similarly, the number of qubits was reduced from $m=8$ to $m=5$ in Fig.~\ref{fig:real-obj}(b).
As shown in Fig.~\ref{fig:real-obj}, the same trend as in Fig.~\ref{fig:int-obj} was observed.
The important aspect here is that almost the same query complexities were achieved despite the reduction in the number of required qubits $m$.
Hence, our proposed real-valued GAS is capable of reducing the size of quantum circuits while maintaining a good performance.

Depending on the channel coefficients and noise, the integer approximation requires a different number of qubits. Since both follow the standard Gaussian distribution, the probability of 0 is the highest, and to deal with smaller values, a larger factor must be multiplied to the objective function, resulting in a larger $m$.
By contrast, direct encoding is capable of keeping $m$ constant.
The only disadvantage is that the probability amplification of $\mathbf{G}^L$ may become insufficient, which was also demonstrated in Fig.~\ref{fig:circuit-real-out}.

\begin{figure}[tb]
	\centering
    \includegraphics[clip, scale=0.6]{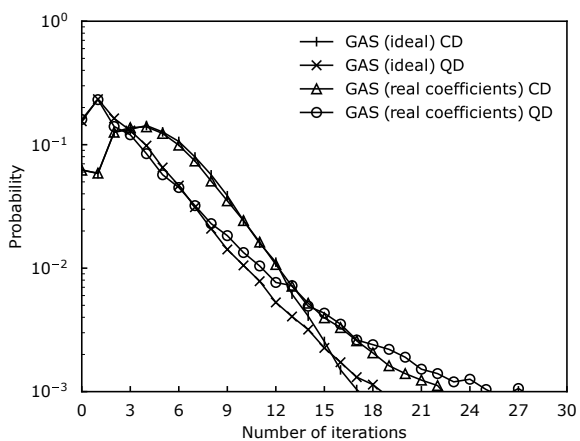}
	\caption{Number of queries required to reach the optimal solution.\label{fig:gasoptcount-qpskran}}
\end{figure}
To investigate the disadvantage of the proposed real-valued GAS and insufficient amplification, in Fig.~\ref{fig:gasoptcount-qpskran}, we generated random channel coefficients and investigated the probability density distribution of the number of queries required to reach the optimal solution, where the parameters were the same as those used in Fig.~\ref{fig:int-obj}(a) except that $m$ was minimized depending on the random channel coefficients.
It was observed in Fig.~\ref{fig:gasoptcount-qpskran} that query complexities in CD and QD increased compared with the ideal case.
Here, the same trend was observed for different SNRs.
Albeit at this expense, the proposed algorithm could reach the optimal solution in any case.
Note that the integer approximation with the same $m$ as in direct encoding could not be plotted in Fig.~\ref{fig:gasoptcount-qpskran} because it was unable to reach the solution in most cases.

\subsection{Effects of Initial Threshold for Further Speedup\label{subsec:initial}}
\begin{figure}[tb]
	\centering
	\subfigure[CD.]{
		\includegraphics[clip, scale=0.6]{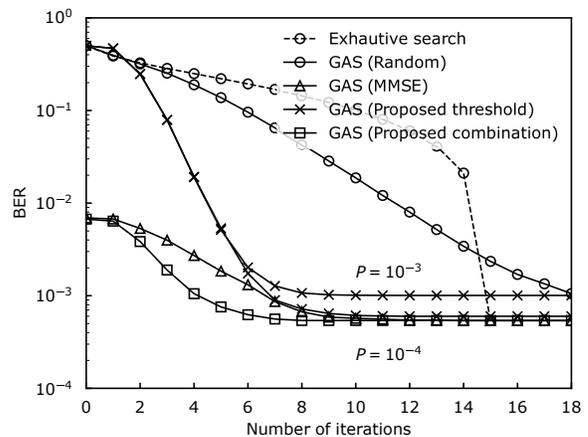}%page reduction
	}
	\subfigure[QD.]{
		\includegraphics[clip, scale=0.6]{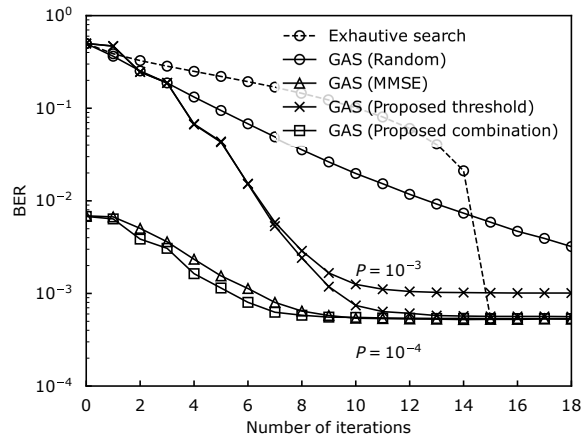}%page reduction
	}
	\caption{BER transition with respect to the number of iterations, where we used random channel coefficients, QPSK, $N_{\mathrm{t}}=2$, $N_{\mathrm{r}}=2$, and $\text{SNR}=20$ dB.\label{fig:iniber-20dB}}
\end{figure}
Finally, in Fig.~\ref{fig:iniber-20dB}, we show the results of evaluating the proposed initial threshold for GAS described in Section~\ref{subsec:mvd}.
Here, we averaged BER with random channel coefficients and noise, considered SNR of $20$ dB, and assumed idealized quantum circuits to examine the impact of the initial threshold only.
Other parameters were the same as those used in Fig.~\ref{fig:real-obj}(a).
Fig.~\ref{fig:iniber-20dB}(a) shows the number of queries in CD, while Fig.~\ref{fig:iniber-20dB}(b) shows these in QD.
Note that the vertical axis is BER rather than the objective function value.
Specifically, at the left end of Fig.~\ref{fig:iniber-20dB}, BER of $0.5$ corresponds to the bit errors between the input bits $\mathbf{b}$ and the random bits $\mathbf{b}_0$, and BER of {$6.8\times 10^{-3}$} corresponds to the errors between $\mathbf{b}$ and the MMSE output $\bar{\mathbf{b}}_0$.
As shown in Fig.~\ref{fig:iniber-20dB}, in both CD and QD, the proposed threshold, $\tilde{y}$, converged to the optimal solution much faster than the classic random threshold.
The slopes in the random and proposed thresholds differed significantly.
This is because the random threshold ranged from the best to the worst cases, resulting in slow convergence in some cases.
To be more specific, in Fig.~\ref{fig:iniber-20dB}, the variance of the random threshold $E(\mathbf{b}_0)$ was $14.30$ at the first iteration.
By contrast, the proposed threshold $\tilde{y}$ is determined by constant factors, $P$ and SNR, and its variance is always zero.
Thus, it significantly improved convergence on average.

It was also found in Fig.~\ref{fig:iniber-20dB} that the proposed threshold achieved the best performance for $P=10^{-4}$ and exhibited lower performance for $P=10^{-3}$.
As described in Section~\ref{subsec:mvd}, $P$ equals the probability that GAS is in a state where the solution is not updated.
That is, an event of $\mathrm{BER}=0.5$ occurred with probability $P=10^{-3}$, and it resulted in the error floor of BER around $10^{-3}$.
This result indicates that the parameter $P$ has no significant impact if it is smaller than BER.
Since the exact BER at a given SNR can be calculated in a closed form in advance, an appropriate $P$ can also be determined in advance accordingly.

Additionally, in Fig.~\ref{fig:iniber-20dB}, the proposed threshold combined with the MMSE output achieved a faster convergence compared with the conventional MMSE only case.
This improvement was greater for CD than for QD.
That is, the proposed threshold is particularly useful for improving the query complexity in CD.
This is because it aims to set a strict threshold even in the case of erroneous MMSE output.
Errors in MMSE estimation lead to higher objective function values and may increase the number of solutions, which can be avoided by adopting the proposed combination.
To be more specific, at the first iteration, the variance of the MMSE-based threshold $\bar{y}$ was $1.58 \cdot 10^{-2}$, while that of the proposed combination $\min(\bar{y}, \tilde{y})$ was much smaller, $3.76 \cdot 10^{-5}$, resulting in the faster convergence.
The performance advantage increased upon increasing SNR, which can be verified from the results shown in Fig.~\ref{fig:cdf_min}.
As confirmed in Fig.~\ref{fig:cdf_min}, the gap between simulated and theoretical values decreased upon increasing SNR.

\begin{figure}[tb]
	\centering
	\subfigure[CD.]{
		\includegraphics[clip, scale=0.6]{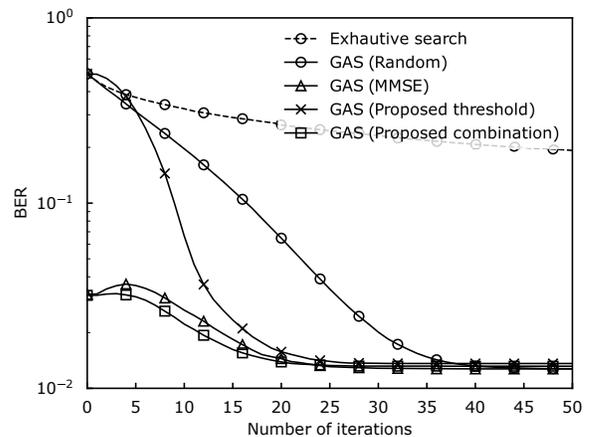}%page reduction
	}
	\subfigure[QD.]{
		\includegraphics[clip, scale=0.6]{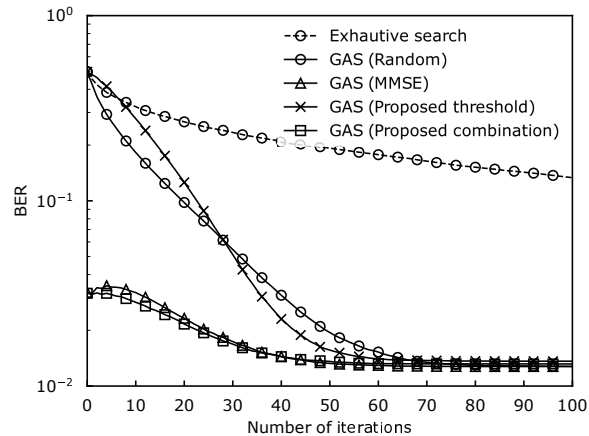}%page reduction
	}
	\caption{BER transition with respect to the number of iterations, where we used random channel coefficients, 16-QAM, $N_{\mathrm{t}}=2, N_{\mathrm{r}}=2$, $\text{SNR}=20$ dB, and $P=10^{-3}$.\label{fig:iniber22qam16-20dB}}
\end{figure}
In Fig.~\ref{fig:iniber-20dB}, the conventional GAS using the random threshold exhibited slower convergence than the classic MLD.
It can be inferred that this slower convergence was caused by the smaller search space.
Since the quadratic speedup is an improvement from $O(2^n)$ to $O(\sqrt{2^n})$, the larger search space leads to the larger reduction.
In Fig.~\ref{fig:iniber22qam16-20dB}, we considered the case of 16-QAM, where other parameters were same as those used in Fig.~\ref{fig:iniber-20dB}.
The classic MLD required $16^2=256$ iterations to reach the optimal solution in the worst case.
As shown in Fig.~\ref{fig:iniber22qam16-20dB}, the conventional MMSE-based threshold exhibited an increase in BER after the first few iterations.
This issue is caused because improvements in objective function values may not correspond to improvements in BER in some cases.
By contrast, the proposed combination successfully avoided the issue and reached the optimal solution with reduced query complexity.
This advantage is expected to increase as the search space size grows, and the proposed approach is especially beneficial for a large-scale MIMO system.

\section{Conclusions and Future Works\label{sec:conc}}
In this paper, we proposed a GAS-based quantum algorithm that supports real-valued HUBO. Then, as an application example, we formulated the MIMO MLD as a HUBO problem. The complexity of MLD exponentially increases with the transmission rate, and low-complexity detectors sacrifice the achievable performance. Unlike in conventional studies, we constructed specific quantum circuits instead of assuming an idealized quantum oracle. This enabled us to analyze the number of qubits and quantum gates in an algebraic manner. To further accelerate the algorithm, we derived the probability distribution of the objective function value and conceived a unique threshold to sample better states. Assuming FTQC, simulations demonstrated the potential for reducing query complexity in CD and providing a quadratic speedup in QD.

Since this paper focused on a specific construction method for quantum circuits and their algebraic analysis, we considered only the hard-decision MLD, instead of error-correcting codes and soft-decision decoding for classical bits, which are common in wireless standards. The error correction capability improves with increasing code distance and length. For example, the maximum code length of 5G NR is 1024 for polar code and 8448 for LDPC. However, with the current computing resources, it is a challenging task to represent such a large-scale system as a specific quantum circuit.
The proposed real-valued GAS can be applied to soft-decision decoding, which will be addressed in our future work.

\section*{Acknowledgement}
IBM, CPLEX and Qiskit are trademarks of International Business Machines Corporation.
The authors are indebted to the Editor and the anonymous reviewers for their invaluable suggestions.

\footnotesize{
	\bibliographystyle{IEEEtran}
	\bibliography{main}

\begin{thebibliography}{10}
\providecommand{\url}[1]{#1}
\csname url@samestyle\endcsname
\providecommand{\newblock}{\relax}
\providecommand{\bibinfo}[2]{#2}
\providecommand{\BIBentrySTDinterwordspacing}{\spaceskip=0pt\relax}
\providecommand{\BIBentryALTinterwordstretchfactor}{4}
\providecommand{\BIBentryALTinterwordspacing}{\spaceskip=\fontdimen2\font plus
\BIBentryALTinterwordstretchfactor\fontdimen3\font minus
  \fontdimen4\font\relax}
\providecommand{\BIBforeignlanguage}[2]{{%
\expandafter\ifx\csname l@#1\endcsname\relax
\typeout{** WARNING: IEEEtran.bst: No hyphenation pattern has been}%
\typeout{** loaded for the language `#1'. Using the pattern for}%
\typeout{** the default language instead.}%
\else
\language=\csname l@#1\endcsname
\fi
#2}}
\providecommand{\BIBdecl}{\relax}
\BIBdecl

\bibitem{nielsen2010quantum}
M.~A. Nielsen and I.~L. Chuang, \emph{Quantum Computation and Quantum
  Information}, 10th~ed.\hskip 1em plus 0.5em minus 0.4em\relax {Cambridge ;
  New York}: {Cambridge University Press}, 2010.

\bibitem{shor1994algorithms}
P.~Shor, ``Algorithms for quantum computation: Discrete logarithms and
  factoring,'' in \emph{Proceedings 35th {{Annual Symposium}} on
  {{Foundations}} of {{Computer Science}}}, Nov. 1994, pp. 124--134.

\bibitem{knuth1976big}
D.~E. Knuth, ``Big {{Omicron}} and big {{Omega}} and big {{Theta}},'' \emph{ACM
  SIGACT News}, vol.~8, no.~2, pp. 18--24, 1976.

\bibitem{grover1996fast}
L.~K. Grover, ``A fast quantum mechanical algorithm for database search,'' in
  \emph{Proceedings of the Twenty-Eighth Annual {{ACM}} Symposium on {{Theory}}
  of Computing}.\hskip 1em plus 0.5em minus 0.4em\relax {Philadelphia,
  Pennsylvania, United States}: {ACM Press}, 1996, pp. 212--219.

\bibitem{bulger2003implementing}
D.~Bulger, W.~Baritompa, and G.~Wood, ``Implementing pure adaptive search with
  {{Grover}}'s quantum algorithm,'' \emph{Journal of Optimization Theory and
  Applications}, vol. 116, pp. 517--529, Mar. 2003.

\bibitem{gidney2018halving}
C.~Gidney, ``Halving the cost of quantum addition,'' \emph{Quantum}, vol.~2,
  p.~74, 2018.

\bibitem{gilliam2020optimizing}
A.~Gilliam, M.~Pistoia, and C.~Gonciulea, ``Optimizing quantum search using a
  generalized version of {{Grover}}'s algorithm,'' \emph{arXiv:2005.06468
  [quant-ph]}, May 2020.

\bibitem{gilliam2021grover}
A.~Gilliam, S.~Woerner, and C.~Gonciulea, ``Grover adaptive search for
  constrained polynomial binary optimization,'' \emph{Quantum}, vol.~5, p. 428,
  Apr. 2021.

\bibitem{kadowaki1998quantum}
T.~Kadowaki and H.~Nishimori, ``Quantum annealing in the transverse {{Ising}}
  model,'' \emph{Phys. Rev. E}, vol.~58, pp. 5355--5363, Nov 1998.

\bibitem{botsinis2013quantum}
P.~Botsinis, S.~X. Ng, and L.~Hanzo, ``Quantum search algorithms, quantum
  wireless, and a low-complexity maximum likelihood iterative quantum
  multi-user detector design,'' \emph{IEEE Access}, vol.~1, pp. 94--122, 2013.

\bibitem{boyer1998tight}
M.~Boyer, G.~Brassard, P.~H{\o}yer, and A.~Tapp, ``Tight bounds on quantum
  searching,'' \emph{Fortschritte der Physik}, vol.~46, no. 4-5, pp. 493--505,
  1998.

\bibitem{durr1999quantum}
C.~Durr and P.~Hoyer, ``A quantum algorithm for finding the minimum,''
  \emph{arXiv:quant-ph/9607014}, Jan. 1999.

\bibitem{botsinis2014fixedcomplexity}
P.~Botsinis, S.~X. Ng, and L.~Hanzo, ``Fixed-complexity quantum-assisted
  multi-user detection for {{CDMA}} and {{SDMA}},'' \emph{IEEE Transactions on
  Communications}, vol.~62, no.~3, pp. 990--1000, Mar. 2014.

\bibitem{botsinis2014lowcomplexity}
P.~Botsinis, D.~Alanis, S.~X. Ng, and L.~Hanzo, ``Low-complexity soft-output
  quantum-assisted multiuser detection for direct-sequence spreading and slow
  subcarrier-hopping aided {{SDMA-OFDM}} systems,'' \emph{IEEE Access}, vol.~2,
  pp. 451--472, 2014.

\bibitem{botsinis2015iterative}
P.~Botsinis, D.~Alanis, Z.~Babar, S.~X. Ng, and L.~Hanzo, ``Iterative
  quantum-assisted multi-user detection for multi-carrier interleave division
  multiple access systems,'' \emph{IEEE Transactions on Communications},
  vol.~63, no.~10, pp. 3713--3727, Oct. 2015.

\bibitem{botsinis2015noncoherent}
------, ``Noncoherent quantum multiple symbol differential detection for
  wireless systems,'' \emph{IEEE Access}, vol.~3, pp. 569--598, 2015.

\bibitem{ye2019quantum}
W.~Ye, W.~Chen, X.~Guo, C.~Sun, and L.~Hanzo, ``Quantum search-aided multi-user
  detection for sparse code multiple access,'' \emph{IEEE Access}, vol.~7, pp.
  52\,804--52\,817, 2019.

\bibitem{alanis2018quantumsearchaided}
D.~Alanis, P.~Botsinis, Z.~Babar, H.~V. Nguyen, D.~Chandra, S.~X. Ng, and
  L.~Hanzo, ``A quantum-search-aided dynamic programming framework for pareto
  optimal routing in wireless multihop networks,'' \emph{IEEE Transactions on
  Communications}, vol.~66, no.~8, pp. 3485--3500, Aug. 2018.

\bibitem{alanis2018quantumaided}
------, ``Quantum-aided multi-objective routing optimization using
  back-tracing-aided dynamic programming,'' \emph{IEEE Transactions on
  Vehicular Technology}, vol.~67, no.~8, pp. 7856--7860, Aug. 2018.

\bibitem{botsinis2017quantumassisted}
P.~Botsinis, D.~Alanis, S.~Feng, Z.~Babar, H.~V. Nguyen, D.~Chandra, S.~X. Ng,
  R.~Zhang, and L.~Hanzo, ``Quantum-assisted indoor localization for uplink
  mm-wave and downlink visible light communication systems,'' \emph{IEEE
  Access}, vol.~5, pp. 23\,327--23\,351, 2017.

\bibitem{botsinis2017coherent}
P.~Botsinis, D.~Alanis, Z.~Babar, S.~X. Ng, and L.~Hanzo, ``Coherent versus
  non-coherent quantum-assisted solutions in wireless systems,'' \emph{IEEE
  Wireless Communications}, vol.~24, no.~6, pp. 144--153, Dec. 2017.

\bibitem{botsinis2019quantum}
P.~Botsinis, D.~Alanis, Z.~Babar, H.~V. Nguyen, D.~Chandra, S.~X. Ng, and
  L.~Hanzo, ``Quantum search algorithms for wireless communications,''
  \emph{IEEE Communications Surveys Tutorials}, vol.~21, no.~2, pp. 1209--1242,
  2019.

\bibitem{fujii2016noise}
K.~Fujii, ``Noise threshold of quantum supremacy,'' \emph{arXiv:1610.03632
  [quant-ph]}, Oct. 2016.

\bibitem{gidney2021how}
C.~Gidney and M.~Eker{\aa}, ``How to factor 2048 bit {{RSA}} integers in 8
  hours using 20 million noisy qubits,'' \emph{Quantum}, vol.~5, p. 433, Apr.
  2021.

\bibitem{ishikawa2021quantum}
N.~Ishikawa, ``Quantum speedup for index modulation,'' \emph{IEEE Access},
  vol.~9, pp. 111\,114--111\,124, 2021.

\bibitem{jay2022ibm}
G.~Jay, ``{{IBM Quantum}} roadmap to build quantum-centric supercomputers,''
  https://research.ibm.com/blog/ibm-quantum-roadmap-2025, May 2022.

\bibitem{stilckfranca2021limitations}
D.~Stilck~Franca and R.~Garcia-Patron, ``Limitations of optimization algorithms
  on noisy quantum devices,'' \emph{Nature Physics}, vol.~17, no.~11, pp.
  1221--1227, 2021.

\bibitem{kim2019leveraging}
M.~Kim, D.~Venturelli, and K.~Jamieson, ``Leveraging quantum annealing for
  large {{MIMO}} processing in centralized radio access networks,''
  \emph{Proceedings of the ACM Special Interest Group on Data Communication},
  pp. 241--255, Aug. 2019.

\bibitem{mondal2021ml}
S.~Mondal, M.~R. Laskar, and A.~K. Dutta, ``{{ML}} criterion based signal
  detection of a {{MIMO-OFDM}} system using quantum and semi-quantum assisted
  modified {{DHA}}/{{BBHT}} search algorithm,'' \emph{IEEE Transactions on
  Vehicular Technology}, vol.~70, no.~2, pp. 1688--1698, Feb. 2021.

\bibitem{babar2020polar}
Z.~Babar, Z.~B. Kaykac~Egilmez, L.~Xiang, D.~Chandra, R.~G. Maunder, S.~X. Ng,
  and L.~Hanzo, ``Polar codes and their quantum-domain counterparts,''
  \emph{IEEE Communications Surveys Tutorials}, vol.~22, no.~1, pp. 123--155,
  2020.

\bibitem{matsumine2019channel}
T.~Matsumine, T.~{Koike-Akino}, and Y.~Wang, ``Channel decoding with quantum
  approximate optimization algorithm,'' in \emph{{IEEE} International Symposium
  on Information Theory}, Jul. 2019, pp. 2574--2578.

\bibitem{ohyama2021intelligent}
T.~Ohyama, Y.~Kawamoto, and N.~Kato, ``Intelligent reflecting surface ({{IRS}})
  allocation scheduling method using combinatorial optimization by quantum
  computing,'' \emph{IEEE Transactions on Emerging Topics in Computing},
  vol.~10, no.~3, 2022.

\bibitem{brassard1998quantum}
G.~Brassard, P.~Hoyer, and A.~Tapp, ``Quantum counting,''
  \emph{arXiv:quant-ph/9805082}, vol. 1443, pp. 820--831, 1998.

\bibitem{brassard2002quantum}
G.~Brassard, P.~Hoyer, M.~Mosca, and A.~Tapp, ``Quantum amplitude amplification
  and estimation,'' \emph{arXiv:quant-ph/0005055}, vol. 305, pp. 53--74, 2002.

\bibitem{3gpp2018ts}
3GPP, ``{{TS}} 138 211 - {{V15}}.2.0 - {{5G}}; {{NR}}; {{Physical}} channels
  and modulation ({{3GPP TS}} 38.211 version 15.2.0 {{Release}} 15),'' 2018.

\bibitem{barenco1995elementary}
A.~Barenco, C.~H. Bennett, R.~Cleve, D.~P. DiVincenzo, N.~Margolus, P.~Shor,
  T.~Sleator, J.~A. Smolin, and H.~Weinfurter, ``Elementary gates for quantum
  computation,'' \emph{Physical Review A}, vol.~52, no.~5, pp. 3457--3467, Nov.
  1995.

\end{thebibliography}
}

\renewenvironment{IEEEbiography}[1]
{\IEEEbiographynophoto{#1}}
{\endIEEEbiographynophoto}

\vspace*{-15pt}
\begin{IEEEbiography}{Masaya~Norimoto}
(S'22) received the B.E. degree from Yokohama National University, Kanagawa, Japan, in 2022. He is currently pursuing the M.E. degree with the Graduate School of Engineering Science, Yokohama National University, Kanagawa, Japan. His research interests include quantum algorithms and wireless communications.
\end{IEEEbiography}
\vspace*{-15pt}
\begin{IEEEbiography}{Ryuhei~Mori}
received the B.E. degree from Tokyo Institute of Technology, Tokyo, Japan in 2008, and the M.Inf. and D.Inf. degrees from Kyoto University, Kyoto, Japan in 2010 and 2013, respectively. From 2013 to 2014, he was a Postdoctoral Fellow at Tokyo Institute of Technology, Tokyo, Japan. He is currently an Assistant Professor of the Department of Mathematical and Computing Sciences, School of Computing, Tokyo Institute of Technology, Tokyo, Japan. His research interests include quantum information, information theory, computer science and statistical physics.
\end{IEEEbiography}
\vspace*{-15pt}
\begin{IEEEbiography}{Naoki~Ishikawa}
(S'13--M'17--SM'22) is an Associate Professor with the Faculty of Engineering, Yokohama National University, Kanagawa, Japan. He received the B.E., M.E., and Ph.D. degrees from the Tokyo University of Agriculture and Technology, Tokyo, Japan, in 2014, 2015, and 2017, respectively. In 2015, he was an academic visitor with the School of Electronics and Computer Science, University of Southampton, UK. From 2016 to 2017, he was a research fellow of the Japan Society for the Promotion of Science. From 2017 to 2020, he was an assistant professor in the Graduate School of Information Sciences, Hiroshima City University, Japan. He was certified as an Exemplary Reviewer of \textsc{IEEE Transactions on Communications} in 2017 and 2021. His research interests include massive MIMO, physical layer security, and quantum speedup for wireless communications.
\end{IEEEbiography}

\end{document}